\documentclass[aps,prd,onecolumn,nofootinbib,superscriptaddress, 11pt]{revtex4} 
\usepackage{hyperref}
\usepackage{graphicx,epstopdf,amsmath,amsfonts,amssymb,appendix,comment} 
\usepackage{color,slashed,subfigure,setspace,footnote,multirow,longtable,braket}
\usepackage{epsfig,mathrsfs,latexsym,color,url,etoolbox}
\usepackage[letterpaper, portrait, margin=0.75in]{geometry}
\definecolor{nicered}{rgb}{0.7,0.1,0.1}
\definecolor{nicegreen}{rgb}{0.1,0.5,0.1}

\usepackage{float}

\hypersetup{colorlinks,citecolor= nicegreen,linkcolor= nicered}

\begin{document}

\singlespacing
\allowdisplaybreaks

{\hfill FERMILAB-PUB-19-146-T, NUHEP-TH/19-04}

\title{Physics with Beam Tau-Neutrino Appearance at DUNE}

\author{Andr\'{e} de Gouv\^{e}a} 
\affiliation{Northwestern University, Department of Physics \& Astronomy, 2145 Sheridan Road, Evanston, IL 60208, USA}
\author{Kevin J. Kelly}
\affiliation{Northwestern University, Department of Physics \& Astronomy, 2145 Sheridan Road, Evanston, IL 60208, USA}
\affiliation{Theoretical Physics Department, Fermi National Accelerator Laboratory, P. O. Box 500, Batavia, IL 60510, USA}
\author{G. V. Stenico}
\affiliation{Northwestern University, Department of Physics \& Astronomy, 2145 Sheridan Road, Evanston, IL 60208, USA}
\affiliation{Instituto de F\'{\i}sica Gleb Wataghin - UNICAMP, 13083-859, Campinas, SP, Brazil}
\author{Pedro Pasquini}
\affiliation{Theoretical Physics Department, Fermi National Accelerator Laboratory, P. O. Box 500, Batavia, IL 60510, USA}
\affiliation{Instituto de F\'{\i}sica Gleb Wataghin - UNICAMP, 13083-859, Campinas, SP, Brazil}

\begin{abstract}
We explore the capabilities of the upcoming Deep Underground Neutrino Experiment (DUNE) to measure $\nu_\tau$ charged-current interactions and the associated oscillation probability $P(\nu_\mu \to \nu_\tau)$ at its far detector, concentrating on how such results can be used to probe neutrino properties and interactions. DUNE has the potential to identify significantly more $\nu_\tau$ events than all existing experiments and can use this data sample to nontrivially test the three-massive-neutrinos paradigm by providing complementary measurements to those from the $\nu_e$ appearance and $\nu_\mu$ disappearance channels. We further discuss the sensitivity of the $\nu_\tau$ appearance channel to several hypotheses for the physics that may lurk beyond the three-massive-neutrinos paradigm: a non-unitary lepton mixing matrix, the $3+1$ light neutrinos hypothesis, and the existence of non-standard neutral-current neutrino interactions. Throughout, we also consider the relative benefits of the proposed high-energy tune of the Long-Baseline Neutrino Facility (LBNF) beam-line.
\end{abstract}


\maketitle

\setcounter{equation}{0}
\section{Introduction}
\label{sec:introduction}

Over the last twenty years, our understanding of neutrino-flavor change as a function of the neutrino proper time has improved exponentially. The old solar and atmospheric neutrino anomalies evolved into the very robust three-massive-neutrinos paradigm, which postulates that neutrinos, while still interacting with ordinary matter as prescribed by the standard model of particle physics (SM), have distinct, non-zero masses and that the neutrino flavor-eigenstates $\nu_{\alpha}$, $\alpha=e,\mu,\tau$, are non-trivial linear superpositions of the neutrino mass-eigenstates $\nu_i$, with masses $m_i$, $i=1,2,3$: $\nu_{\alpha}=U_{\alpha i}\nu_i$, where $U_{\alpha i}$ are the coefficients of the $3\times 3$ unitary lepton mixing matrix $U$. 

Assuming the three-massive-neutrinos paradigm is correct, the oscillation parameters -- those that determine $U$, along with the neutrino mass-squared differences $\Delta m^2_{ij} \equiv m_i^2-m_j^2$, $i,j=1,2,3$ -- are well constrained by existing neutrino data, with a few exceptions, including the neutrino mass ordering (or the sign of $\Delta m^2_{31}$) and the strength of leptonic CP-invariance violation. When it comes to both confirming the three-massive-neutrinos paradigm and determining the oscillation parameters, all current statistical power comes from studies of $\nu_{\mu}$ and $\nu_e$ (plus antineutrinos) disappearance and $\nu_{\mu}\to\nu_e$ (plus antineutrinos) appearance. While there is definitive evidence for $\nu_{\mu}\to\nu_{\tau}$ appearance from both atmospheric \cite{Abe:2012jj,Li:2017dbe,Aartsen:2019tjl} and beam \cite{Agafonova:2015jxn} neutrino experiments, the quantitative impact of the current $\nu_{\tau}$ appearance data is, at best, sub-dominant.\footnote{A study of the new-physics reach of the OPERA data was recently made available in the preprint archives \cite{Meloni:2019pse} and the collaboration made its latest analysis of the oscillation hypothesis available in Ref.~\cite{Agafonova:2019npf}.}

Among the goals of the next-generation long-baseline neutrino experiments -- the Long-Baseline Neutrino facility to the Deep Underground Neutrino Experiment (LBNF-DUNE) in the United States and the Tokai to Hyper-Kamiokande (T2HK) experiment in Japan -- are the precision measurements of neutrino oscillation parameters, exploring CP-invariance violation in the neutrino sector, and, perhaps most important, testing the validity of the three-massive-neutrinos paradigm and looking for more new physics in the neutrino sector. There are several studies in the literature, including those pursued by the collaborations, of the physics reach of LBNF-DUNE and T2HK, including their sensitivities to a variety of hypothetical new neutrino physics scenarios. All of these explore the $\nu_{\mu}$ disappearance and the $\nu_{\mu}\to\nu_e$ appearance (plus antineutrinos) channels. 

Here, instead, we explore the physics reach of the beam $\nu_{\tau}$ appearance data that is accessible to LBNF-DUNE. Direct measurements of $\nu_{\tau}$ appearance, both in atmospheric and beam neutrino experiments, are very challenging, for a variety of reasons. All neutrino experiments are of the fixed-target type and the large $\tau$ mass translates into relatively large thresholds for charged-current $\nu_{\tau}$ scattering off of ordinary matter ($E_{\nu_{\tau}}\gtrsim 3.35$~GeV for $\nu_{\tau}+N\to\tau+N$, where $N$ is a nucleon, and $E_{\nu_{\tau}}\gtrsim3.1$~TeV for $\nu_{\tau}+e\to\tau+\nu_e$). Given the beam energies of LBNF-DUNE and T2HK, driven by the requirement that large $\Delta m_{31}^2$-driven oscillation effects are observed in the far detector, only LBNF-DUNE is expected to observe beam events with neutrino energies above the $\tau$-threshold. Even at LBNF-DUNE, phase-space-suppression effects are large and the $\nu_{\tau}$ appearance event sample is expected to contain, for the lifetime of the experiment, between 100 and 1,000 events. Identifying and reconstructing $\tau$-leptons in neutrino detectors is also very challenging. The tracking resolution of liquid argon detectors, around several millimeters,  is such that $\tau$-leptons decay promptly and hence must be identified via their decay products. Furthermore, all $\tau$ leptons decay channels involve missing energy -- $\tau\to\nu_{\tau}~+$~something else. These imply that the $\nu_{\tau}$ appearance channel is ``dirty'' and that the reconstruction of the $\nu_{\tau}$ energy is a larger challenge than that of the $\nu_{\mu}$ and $\nu_e$  energies. We discuss our simulations of the $\nu_{\tau}$ sample at LBNF-DUNE in Sec.~\ref{sec:DUNE}, including background estimates and the challenges of neutrino-energy reconstruction. 

In spite of all the challenges, LBNF-DUNE is expected to collect an unprecedented number of reconstructed beam $\nu_{\tau}$ events. Here, we proceed to understand what nontrivial particle physics information one should be able to extract. We explore, assuming the three-massive-neutrinos paradigm, how well one can measure the $\nu_{\tau}$ charged-current scattering cross section, and, assuming the standard model expectation for the $\nu_{\tau}$ charged-current scattering cross section, how well the $\nu_{\tau}$ appearance data can constrain oscillation parameters. As far as the latter exercise is concerned, we compare, in different ways, the sensitivity of the $\nu_{\tau}$ appearance sample with that of the $\nu_{e}$ appearance and $\nu_{\mu}$ disappearance samples. These results are presented and discussed in detail in Sec.~\ref{sec:3Nu}.

While, not surprisingly, the reach of the $\nu_{\tau}$ appearance sample is comparatively weak, the value of using different oscillation channels to measure the oscillation parameters cannot be overstated. Indeed, comparing the results of different oscillation channels consists of one of the most robust tests of the three-massive-neutrinos paradigm.  In order to further pursue how the $\nu_{\tau}$ appearance sample complements the search of physics beyond the three-massive-neutrinos paradigm, we explore different concrete scenarios, including the existence of new heavy and light neutrino degrees of freedom -- Sec.~\ref{sec:nonunitary} and Sec.~\ref{sec:sterile}, respectively -- and the existence of new neutral--current neutrino--matter interactions (Sec.~\ref{sec:NSI}).

In Sec.~\ref{sec:conclusion}, we present some concluding remarks and highlight a few future research directions one can pursue with $\nu_{\tau}$-appearance at LBNF-DUNE, including searches in the near detector facility and the atmospheric $\nu_{\tau}$ sample. 

\setcounter{equation}{0}
\section{The $\nu_{\tau}$ Sample at DUNE}
\label{sec:DUNE}

The production of $\tau$ leptons by charged-current $\nu_{\tau}$--nucleus scattering requires neutrino energies $E_\nu \gtrsim 3.4$~GeV. Furthermore, the prompt decay of the $\tau$ combined with the fact that all $\tau$ decays contain $\nu_{\tau}$ in the final state prove a challenge for both identifying a scattering event as a $\nu_\tau$ charged-current interaction and reconstructing the initial neutrino energy. Nonetheless, the LBNF beam will have a significant portion of its flux above the $\tau$ production energy threshold and liquid argon detectors are excellent at reconstructing final-state particles, allowing for at least a modest but unique  $\nu_{\tau}$ data sample. Here, we discuss our expectations regarding the capability of DUNE to identify $\nu_{\tau}$ interactions and measure their energies.

We assume that the DUNE far detector, 1300 km from the neutrino source at Fermilab, consists of 40~kton of liquid argon (fiducial mass) and that the LBNF beam will deliver $1.1 \times 10^{21}$ protons on target (POT) per year. We consider three different modes of beam operation -- forward horn current (which we will refer to as ``neutrino mode''), reverse horn current (``antineutrino mode''), and the currently-under-consideration tau-optimized flux~\cite{LauraFlux} (``high-energy mode''). The high-energy mode contains significantly higher-energy neutrinos than the other two modes, with a large fraction of the neutrinos above the $\tau$ threshold. For neutrino and antineutrino modes, we simulate event yields using the ``CP-optimized fluxes'' from Refs.~\cite{LauraFlux,Acciarri:2015uup}. In our analyses we consider two different hypotheses for DUNE's data-collection strategy. We (a) assume $3.5$ years of operation each in neutrino and antineutrino modes, for total of seven years. We do not perform any analyses considering only the high-energy mode, but (b) we perform analyses combining $3 + 3$ years of neutrino and antineutrino mode with an additional year of data taking in the high-energy mode, also for a total of seven years. We refer to these two data collection strategies as ``$3.5+3.5$'' and ``$3+3+1$'', respectively.

When a $\nu_\tau$ interacts via a charged-current interaction, producing a $\tau$ lepton, the decay length of the $\tau$ is significantly smaller than the resolution of the DUNE detector, meaning that one must reconstruct the $\tau$ decay products in order to classify the incoming  neutrino as a $\nu_\tau$. The authors of Ref.~\cite{Conrad:2010mh} investigated the capability of liquid argon detectors to identify $\nu_\tau$ events using hadronic $\tau$ decays, which also make up the largest branching fraction for the $\tau$ (around 65\% \cite{Tanabashi:2018oca}). This approach takes advantage of the detector's capability to identify pions and kaons and to measure their kinematic properties. Understanding the usefulness of the leptonic $\tau$ decays, which are, naively, heavily contaminated by charged-current $\nu_{\mu}$ or $\nu_e$ scattering, is beyond the scope of this manuscript. Building on Ref.~\cite{Conrad:2010mh}, efforts within the DUNE collaboration~\cite{AdamSlides} have optimistically estimated that one can isolate a $\nu_\tau$-rich event sample where 30\% of hadronically-decaying $\tau$ events are successfully identified while only 0.5\% of neutral-current background events remain. More sophisticated analyses are currently under intense investigation by the collaboration. We restrict ourselves to this rather optimistic scenario throughout this work, and hope that the results discussed here will help motivate more detailed studies of the $\nu_{\tau}$-reconstruction capabilities of DUNE and other liquid argon detectors. 

\begin{figure}[ht]
\centering
\includegraphics[width=0.6\linewidth]{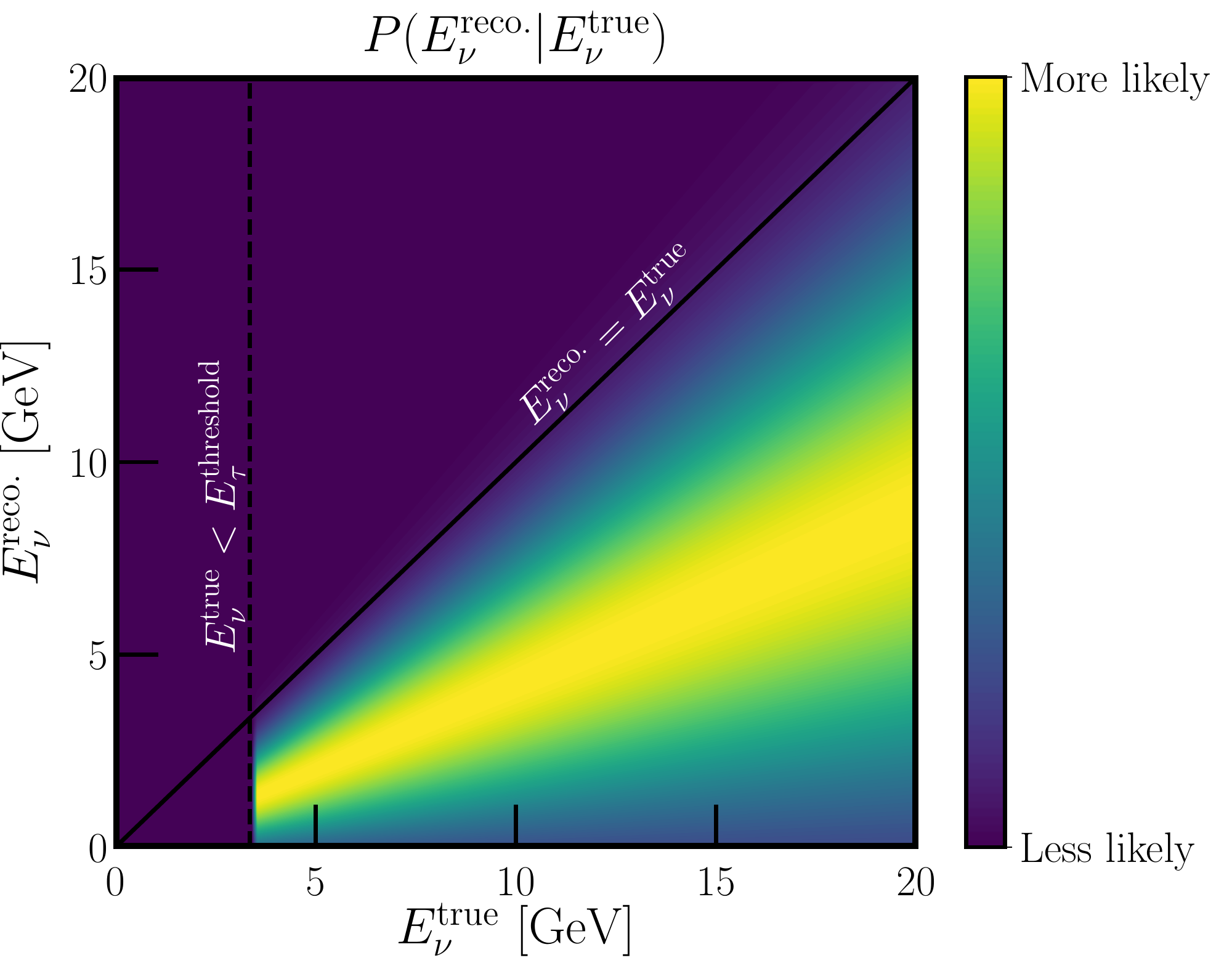}
\caption{Migration matrix for hadronically-decaying $\tau$ leptons produced via $\nu_\tau$ charged-current interactions. The assumed bias is $45\%$ and the resolution is $25\%$, see text for details. No migration exists below $E_\nu^{\mathrm{true}} \approx 3.4$~GeV, below which the scattering process is kinematically forbidden.}
\label{fig:MM}
\end{figure}

When performing an oscillation analysis, it is paramount to measure the energy of the incoming neutrino. This is relatively straightforward\footnote{Albeit not at all trivial and still under intense investigation, see, for example, Refs.~\cite{DeRomeri:2016qwo,Friedland:2018vry}.} for charged-current events induced by $\nu_e$ and $\nu_\mu$, where the outgoing charged lepton is well-reconstructed. However, $\tau$ leptons are impossible to reconstruct perfectly, as their decays necessarily involve at least one outgoing neutrino. We have constructed a migration matrix mapping the true neutrino energy $E_\nu^\mathrm{true}$ to the reconstructed neutrino energy $E_\nu^\mathrm{reco}$, parameterizing it with a bias in the preferred reconstructed energy as a function of the true energy, as well as an uncertainty on the reconstructed energy. This is done by simulating final-state hadronic $\tau$-decays using {\sc MadGraph}~\cite{Alwall:2014hca} for different $\tau$ energies and computing the assumed-to-be observed energy in the hadronic system. Our simulations are consistent with the following simple picture. For a given $E_\nu^\mathrm{true}$, the reconstructed energy follows a Gaussian distribution with a mean value $bE_\nu^\mathrm{true}$ and width $\sigma_E = rE_\nu^\mathrm{true}$, where $b$ is the bias and $r$ is the resolution of the measurement. Our simulations point to $b \approx 45\%$ and $r \approx 25\%$. While the bias does not have a significant impact on the quantitative results presented in the following section, the resolution does; more detailed studies indicate that the DUNE collaboration will be able to achieve $8\% \lesssim r \lesssim 25\%$, and we choose the conservative upper limit~\cite{AdamComm}. Fig.~\ref{fig:MM} depicts the migration matrix we have obtained and will use for all forthcoming analyses. We have also performed a more realistic simulation using the {\sc Genie3.0} software~\cite{Andreopoulos:2015wxa}, and the results we obtained agree with Fig.~\ref{fig:MM}.

\begin{figure}[ht]
\centerline{
\includegraphics[width=1\textwidth]{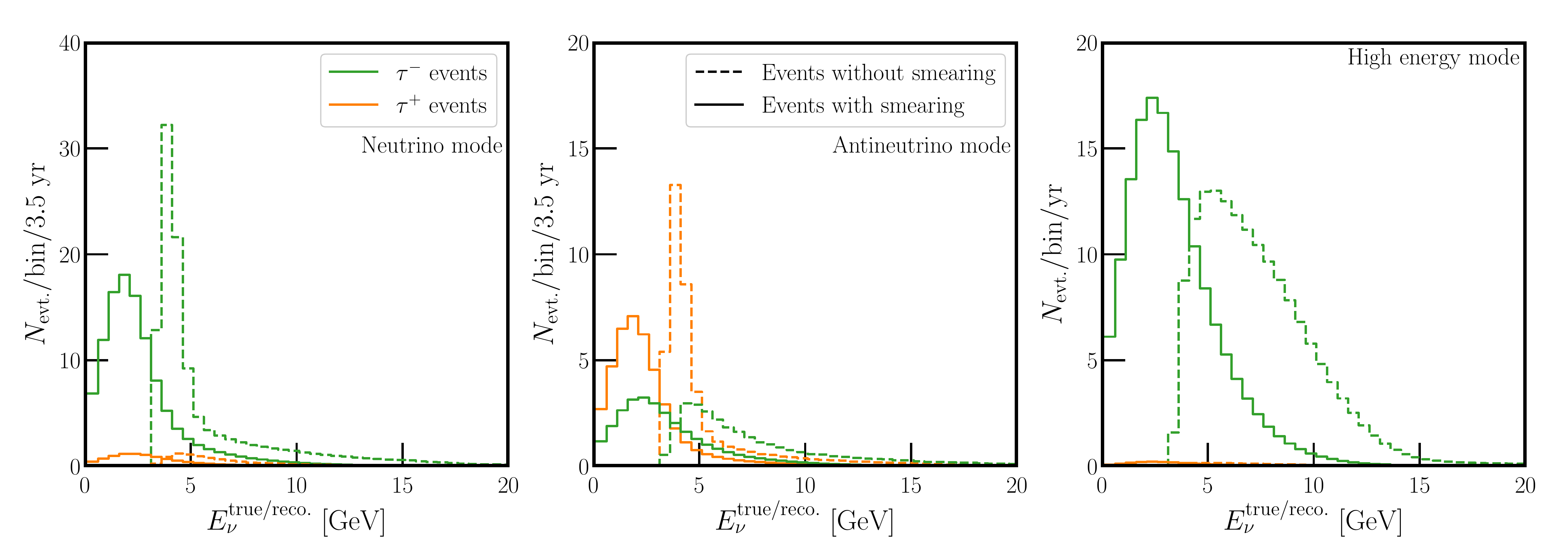}}
\caption{Expected number of $\nu_\tau$-identified signal events per 0.5 GeV bin as a function of the true (dashed) or reconstructed (solid) neutrino energy. The left panel displays the expected number of events when in neutrino mode, the center panel displays the antineutrino mode, and the right panel displays high-energy mode events. In each panel, we show the contribution due to neutrinos in green and antineutrinos in orange. Each distribution has been normalized to the expected runtime in each mode, $3.5$ years for neutrino and antineutrino modes and $1$ year for high-energy mode.}\label{fig:EventYields}
\end{figure}

Virtually no $\nu_{\tau}$ are produced at the neutrino source. In the DUNE detector, however, the three-massive-neutrinos paradigm predicts a healthy $\nu_{\tau}$ flux, mostly from $\nu_{\mu}\to\nu_{\tau}$  oscillations over the 1300~km baseline. Assuming the three-massive-neutrinos paradigm, the SM $\nu_{\tau}$ scattering cross section, and incorporating the expected identification capability for hadronic $\tau$ events, we calculate the expected number of reconstructed events as a function of neutrino energy for each mode of beam operation. Fig.~\ref{fig:EventYields} depicts the expected number of signal events for the neutrino mode (left), the antineutrino mode (center), and the high-energy mode (right). Here, we assumed the following values for the oscillation parameters \cite{Tanabashi:2018oca}, in agreement with the most recent global-fit results obtained by the NuFit collaboration~\cite{Esteban:2018azc}: 
\begin{eqnarray}
&\sin^2\theta_{12} = 0.310,~~\sin^2\theta_{13} = 0.02240,~~\sin^2\theta_{23} = 0.582,~~\delta_{CP} = 217^\circ = -2.50~{\rm rad}, \nonumber \\ 
& \Delta m_{21}^2 = 7.39 \times 10^{-5}~{\rm eV^2},~~\Delta m_{31}^2 = +2.525\times 10^{-3}~ {\rm eV^2}. \label{eq:nufit}
\end{eqnarray}
 Unless otherwise stated, we will assume these oscillation parameters to be the true values in our analyses going forward. We discuss the details on the $\nu_{\mu}\to\nu_{\tau}$ oscillation probability in the next section. Fig.~\ref{fig:EventYields} depicts both ``smeared'' (solid histograms) and ``unsmeared'' (dashed histograms) event yields using the energy migration matrix discussed above. We divide the simulated data in energy bins of constant width $\Delta E_\nu = 0.5$~GeV, between $0$ and $20$ GeV, for our analyses.

\begin{figure}[ht]
\centerline{
\includegraphics[width=1\textwidth]{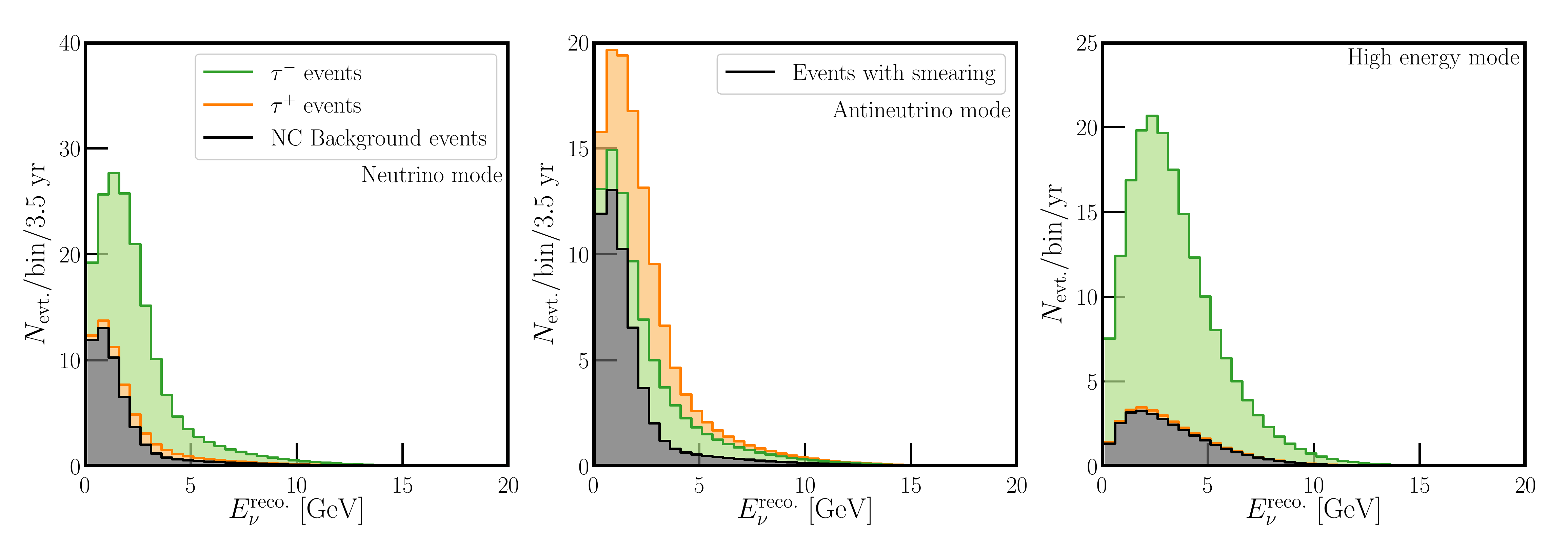}}
\caption{Expected number of reconstructed $\nu_\tau$ events per 0.5 GeV bin as a function of reconstructed energy. The left panel displays the expected number of events when in neutrino mode, the center panel displays the antineutrino mode, and the right panel displays high-energy mode events. We display stacked histograms of background (grey), events from $\overline{\nu}_\tau$ producing $\tau^+$ leptons (orange), and events from $\nu_\tau$ producing $\tau^-$ (green). Each distribution has been normalized to the expected runtime in each mode, $3.5$ years for neutrino and antineutrino modes and $1$ year for high-energy mode.}\label{fig:EventYields_Stack}
\end{figure}

Fig.~\ref{fig:EventYields_Stack} depicts the expected event yields -- stacked -- taking neutral-current backgrounds into account, along with the energy smearing. In seven years of running, we anticipate a healthy, relatively clean $\nu_{\tau}$-appearance sample. In the $3.5+3.5$ case, the sample includes over 200~events, while in the $3+3+1$ running scheme we expect over 300~events. In both cases, neutral-current backgrounds make up less than 50\% of all events~\cite{AdamSlides}.

From time to time, we will compare the results obtained using our simulated $\nu_\tau$ data sample with those obtained using samples of $\nu_\mu \to \nu_\mu$ disappearance and $\nu_\mu \to \nu_e$ appearance (and their CP-conjugated channels). The details of those simulations are provided in Refs.~\cite{Berryman:2015nua,deGouvea:2015ndi,Berryman:2016szd,deGouvea:2016pom,deGouvea:2017yvn} with fluxes updated to match those in Refs.~\cite{LauraFlux,Acciarri:2015uup}. The expected yields are presented in Appendix~\ref{appendix:others}.

In our analyses, we include a 25\% normalization uncertainty on the number of signal events, a conservative assumption related to systematic uncertainties regarding the neutrino flux, cross section, etc. We find, in practice, that this uncertainty does not have a strong impact on the results discussed in the next section. 

Assuming the three-massive-neutrinos paradigm is correct, one can use the $\nu_{\tau}$ sample to measure the $\nu_{\tau}$ charged-current cross section on argon. Excluding the 25\% normalization uncertainty mentioned in the last paragraph\footnote{When measuring the charged-current cross section, one should not include systematic effects related to the cross section. Here, for simplicity, we are assuming that other normalization-related uncertainties will be constrained in a variety of ways, including measurements of $\nu_{\mu}$ disappearance and $\nu_e$ appearance.} and holding oscillation parameters fixed, we find that DUNE can measure the flux-averaged $\nu_{\tau}$ charged-current cross section on argon to be $\langle \sigma \rangle = \langle \sigma_0 \rangle(1 \pm 0.07)$ (or $\langle \sigma \rangle = \langle \sigma_0 \rangle(1^{+0.15}_{-0.14})$ at the two sigma level), where $ \langle \sigma_0 \rangle$ is the SM expectation, after seven years of data taking (3.5 years in neutrino and antineutrino mode, respectively). This can be compared to current results from OPERA, $\langle \sigma \rangle = \langle \sigma_0 \rangle(1.2^{+0.6}_{-0.5})$~\cite{Agafonova:2018auq} and Super-Kamiokande, $\langle \sigma \rangle = \langle \sigma_0 \rangle(1.47 \pm 0.32)$~\cite{Li:2017dbe}. Note that these are not apples-to-apples comparisons. The expected flux-averaged neutrino energy for DUNE is $\langle E_\nu \rangle \simeq 4$~GeV, compared to the one for OPERA $\langle E_\nu \rangle \simeq 10$~GeV and Super-Kamiokande $\langle E_\nu \rangle \simeq 7$~GeV.

\setcounter{equation}{0}
\section{Physics with the beam $\nu_{\tau}$ sample}

In this section, we simulate and analyze the $\nu_{\tau}$ sample that is expected to be collected at DUNE after either 3.5 years of running in the neutrino mode plus 3.5 years of running in the antineutrino mode ($3.5+3.5$) or 3 years of running in both the neutrino and antineutrino modes plus 1 year of running in the high energy mode ($3+1+1$), as discussed in the previous section. The simulated data are consistent with the three-massive-neutrinos paradigm, where the input oscillation parameters are set to the values listed in Eq.~(\ref{eq:nufit}); the associated event yields for $\nu_{\tau}$-appearance are depicted in Figs.~\ref{fig:EventYields} and~\ref{fig:EventYields_Stack} while the ones for disappearance and $\nu_e$-appearance are depicted in Appendix~\ref{appendix:others}. When discussing oscillation parameters, following tradition, we will refer to $\Delta m^2_{21}$ and $\sin^2\theta_{12}$ as the `solar parameters,' while $\Delta m^2_{31}$ and $\sin^2\theta_{23}$ are referred to as the `atmospheric parameters.'  

We will discuss four distinct hypotheses: the three-massive-neutrinos paradigm (Sec.~\ref{sec:3Nu}), a non-unitary lepton mixing matrix, in the context of heavy\footnote{For the purpose of neutrino beam experiments, ``heavy'' refers to new neutrino states heavier than the kaon and hence not kinematically accessible in pion or kaon decay. ``Light,'' on the other hand, refers to new neutrino states much lighter than the pion, so there is no relative phase-space suppression for the production of the new neutrino mass eigenstate in meson decays.} new neutrino states (Sec.~\ref{sec:nonunitary}), the existence of a fourth, light neutrino (Sec.~\ref{sec:sterile}), and non-standard neutral-current neutrino--matter interactions (Sec.~\ref{sec:NSI}). We do not advocate for or favor any particular hypothesis for the physics that may exist beyond the three-massive-neutrinos paradigm. Instead, these are to be viewed as different, quantifiable modifications to the neutrino oscillation probabilities that serve as proxies for what may ultimately turn out to be the new physics. All hypotheses have been heavily scrutinized in the past so it is easy to make comparisons between different neutrino oscillation experiments and different oscillation channels.

\subsection{Three-Massive-Neutrinos Paradigm}
\label{sec:3Nu}

Assuming the three-massive-neutrinos paradigm, in vacuum, for the LBNF-DUNE baseline $L=1300$~km and neutrino energies above $\tau$ threshold ($E_{\nu}\gtrsim 3.4$~GeV), 
\begin{equation}
P(\nu_\mu \to \nu_\tau) = 4|U_{\mu 3}|^2 |U_{\tau 3}|^2 \sin^2{\left(\frac{\Delta m_{31}^2 L}{4 E_\nu}\right)} + {\rm subleading}.
\label{eq:pmutau}
\end{equation}
Numerically
\begin{equation}
\frac{\Delta m_{31}^2 L}{4 E_\nu}=0.75\left(\frac{\pi}{2}\right)\times \left(\frac{3.5\rm GeV}{E_{\nu}}\right)\left(\frac{\Delta m^2_{31}}{2.5\times 10^{-3}~\rm eV^2}\right)\left(\frac{L}{1300~\rm km}\right),
\label{eq:phase}
\end{equation}
such that all reconstructed $\nu_{\tau}$ appearance events occur above the first oscillation maximum. The oscillation amplitude is $4|U_{\tau3}|^2|U_{\mu3}|^2=\cos^4\theta_{13}\sin^22\theta_{23}\sim 0.95$, using the NuFit results listed in Eqs.~(\ref{eq:nufit}). The subleading terms include the ``solar'' oscillations and the interference term. Here, the subleading terms are indeed subleading: $\Delta m^2_{21}/\Delta m^2_{31}\sim 0.03$ and, unlike the case of $\nu_\mu \to \nu_e$ oscillations, all relevant elements of the lepton mixing matrix are large.\footnote{In the case of $\nu_\mu \to \nu_e$ oscillations, the leading term is proportional to $|U_{e3}|^2\sim 0.022\ll 1$. In this case, the interference term can be almost as large as the leading term for LBNF-DUNE neutrino energies and baseline.} More quantitatively, for the energies of interest, we expect the interference term to be around a few percent of the leading term while the solar term is at the per-mille level. Matter effects, which we include in all numerical computations, are small and do not modify this picture. We provide more details in Appendix~\ref{appendix:prob}.

In summary, as far as relevant DUNE energies and baseline are concerned, $P(\nu_\mu \to \nu_\tau)$ depends, at leading order, only on two effective oscillation parameters: $\Delta m^2_{31}$ and 
\begin{equation}
\sin^22\theta_{\mu\tau}\equiv 4|U_{\mu 3}|^2 |U_{\tau 3}|^2.
\label{eq:sin2theta}
\end{equation}
Furthermore, there is no access to even the first oscillation maximum. Under these circumstances, it is virtually impossible, using $\nu_\mu \to \nu_\tau$, to determine the octant of $\theta_{23}$ (whether $\theta_{23} < \pi/4$ or $\theta_{23} > \pi/4$), since $P(\nu_\mu \to \nu_\tau)$ is, at leading order, invariant upon $|U_{\mu3}|^2\leftrightarrow|U_{\tau3}|^2$. The same is true for the neutrino mass ordering since resolving the neutrino mass ordering with neutrino oscillations requires visible matter effects or the ability to resolve more than one mass-squared difference.  

\begin{figure}[ht]
\centering
\includegraphics[width=0.65\textwidth]{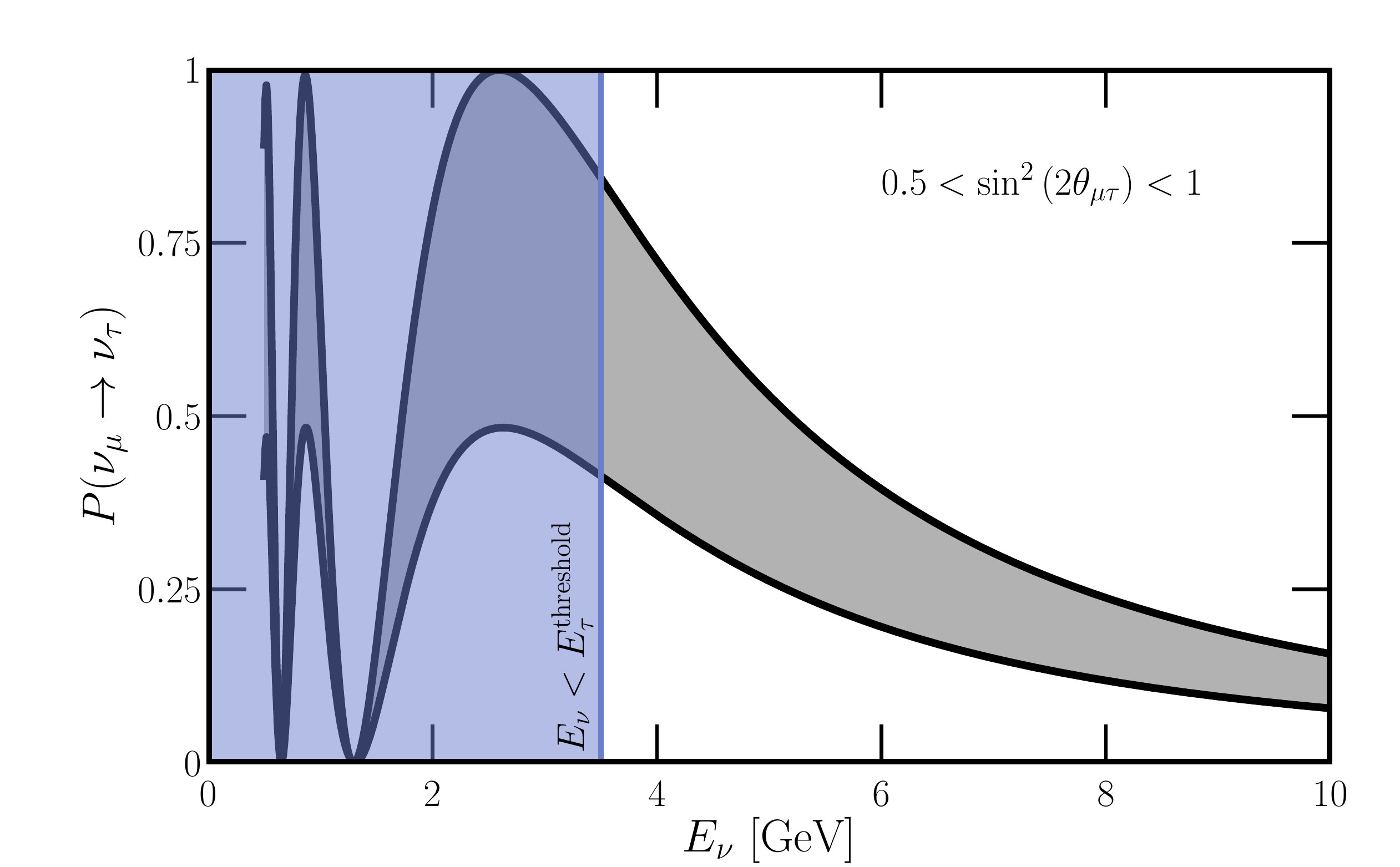}
\caption{The oscillation probability $P(\nu_\mu \to \nu_\tau)$ as a function of neutrino energy $E_\nu$. The blue-shaded region corresponds to neutrino energies below the $\tau$ production threshold. The grey band is spanned by allowing values of  $\sin^2 2\theta_{\mu\tau} \in [0.5,1]$. All other oscillation parameters are fixed to the values listed in Eqs.~(\ref{eq:nufit}).}
\label{fig:osci_prob} 
\end{figure}

Fig.~\ref{fig:osci_prob} depicts the $\nu_{\mu}\to\nu_{\tau}$ oscillation probability. All oscillation parameters are fixed to the values listed in Eqs.~(\ref{eq:nufit}), except for $\sin^2\theta_{23}$. The shaded region corresponds to $\sin^22\theta_{\mu\tau} \in [0.5,1]$. 

\begin{figure}[ht]
\centerline{
\includegraphics[width=1\textwidth]{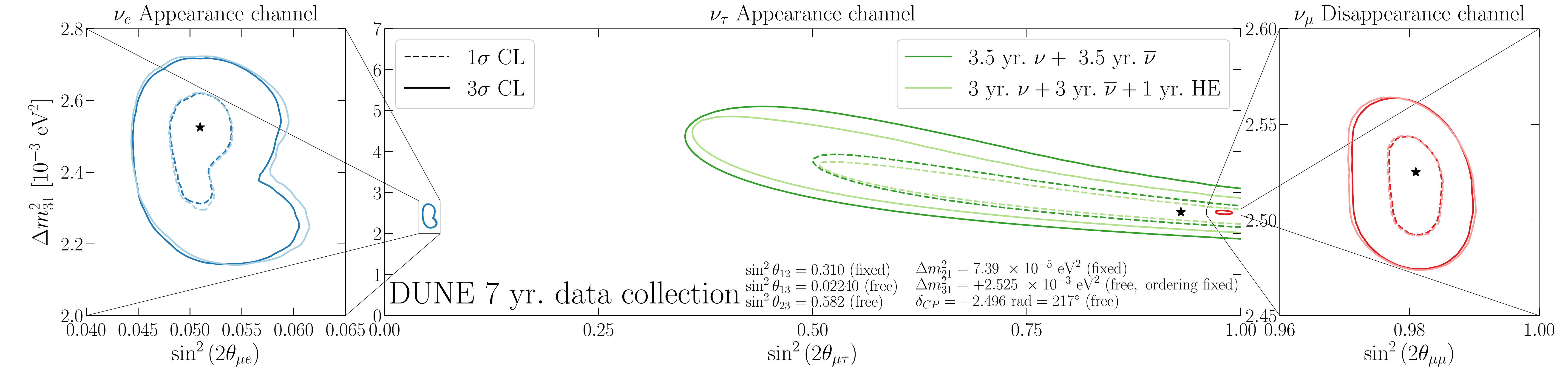}}
\caption{Expected measurement potential of seven years of data collection at DUNE, assuming separate analyses of the three oscillation channels: $\nu_e$ Appearance (left), $\nu_\tau$ Appearance (center), and $\nu_\mu$ Disappearance (right). We show $1\sigma$ (dashed lines) and $3\sigma$ (solid) CL regions of the measurement of the dominant effective mixing angle in each channel $\sin^2 (2\theta_{\mu \beta})$ (see Eqs.~(\ref{eq:sin2theta}) and (\ref{eq:MixingAppDis}) for definitions) and the mass-squared difference $\Delta m_{31}^2$. The $\nu_e$ and $\nu_\mu$ $3\sigma$ results are shown as inserts in the center panel. In these analyses, we fix $\sin^2\theta_{12}$ and $\Delta m_{21}^2$ to their best-fit values, fix the neutrino mass ordering to normal, and marginalize over $\sin^2\theta_{13}$ and $\delta_{CP}$. In each panel, we show results assuming $3.5$ years each of data collection in neutrino and antineutrino modes in dark colors, and results assuming $3$ years in neutrino/antineutrino modes and $1$ year in high-energy mode in light colors.}
\label{fig:two_param}
\end{figure}

We analyze the simulated $3.5+3.5$ $\nu_{\tau}$ appearance data. Fig.~\ref{fig:two_param}(center) depicts the preferred regions of the  $\sin^22\theta_{\mu\tau}\times \Delta m^2_{31}$-plane at the one (solid) and three (dashed) sigma level. The star indicates the best-fit point. In the analysis, the solar oscillation parameters are held fixed, along with the neutrino mass ordering (normal), while the other mixing parameters orthogonal to $\sin^2(2\theta_{\mu\tau})$ -- in practice, $\sin^2\theta_{13}$ and $\delta_{CP}$ -- are marginalized. We do not include any other external information on the mixing parameters, including the current very precise measurement of $\sin^2\theta_{13}$ by reactor neutrino experiments. We emphasize that only the $\nu_{\tau}$ appearance data from DUNE is used in the fit -- the $\nu_{\mu}$ disappearance and the $\nu_e$ appearance data are not included in this fit -- so this is, except for the solar parameters, a $\nu_\tau$-appearance-only measurement of neutrino oscillation parameters. 

This result is easy to understand. Since the accessible neutrino energies are all above the first oscillation maximum, the $\nu_{\tau}$ data, to zeroth order, measures the combination $\sin^22\theta_{\mu\tau}\times (\Delta m^2_{31})^2$. This behavior cuts-off at $\Delta m^2_{31}\sim 5\times 10^{-3}$~eV$^2$, when the oscillation maximum shifts significantly to energy values above $\tau$-threshold. 

We perform similar analyses of the simulated $\nu_e$-appearance and $\nu_{\mu}$-disappearance samples. The results are depicted in Fig.~\ref{fig:two_param}(left) and Fig.~\ref{fig:two_param}(right), respectively. For the sake of comparison, we include these preferred regions in Fig.~\ref{fig:two_param}(center), making it clear that they correspond to different observables in the horizontal axis. The effective mixing angles are generalizations of Eq.~(\ref{eq:sin2theta}),
\begin{equation}\label{eq:MixingAppDis}
\sin^22\theta_{\mu e}\equiv 4|U_{\mu3}|^2 |U_{e3}|^2,~~~~~~\sin^22\theta_{\mu\mu}\equiv 4|U_{\mu3}|^2(1-|U_{\mu3}|^2).
\end{equation}
In both analyses the solar parameters and the neutrino mass ordering are held fixed, and we marginalize over all other orthogonal oscillation parameters. As with the $\nu_{\tau}$-appearance analysis, we do not include other external information on the mixing parameters.

Fig.~\ref{fig:two_param} reveals that, in some sense, and as expected, the $\nu_{\tau}$-appearance sample is much less precise than the $\nu_{e}$-appearance or $\nu_{\mu}$-disappearance samples. On the other hand, the three data sets are virtually independent and, strictly speaking, measure different phenomena. The information they provide is complementary. Assuming the three-massive-neutrinos paradigm, however, the different $\sin^22\theta_{\mu\alpha}$, $\alpha=e,\mu,\tau$ are not independent but satisfy $\sin^22\theta_{\mu e}+\sin^22\theta_{\mu \tau}=\sin^22\theta_{\mu\mu}$. This assumption would allow one to determine $\sin^22\theta_{\mu \tau}$ at (better than) the percent level with DUNE $\nu_{e}$-appearance and $\nu_{\mu}$-disappearance data only. We return to tests of the unitarity of the mixing matrix in the next subsection. Fig.~\ref{fig:two_param} also depicts the results obtained in the $3+3+1$ case (faint colors). These are virtually indistinguishable from those obtained in the $3.5+3.5$ case (bold colors). In the $\nu_{\tau}$-appearance channel, the larger sample size leads to marginally tighter allowed regions than in the $3.5+3.5$ case.

\begin{figure}[ht]
\centering
\includegraphics[width=0.85\textwidth]{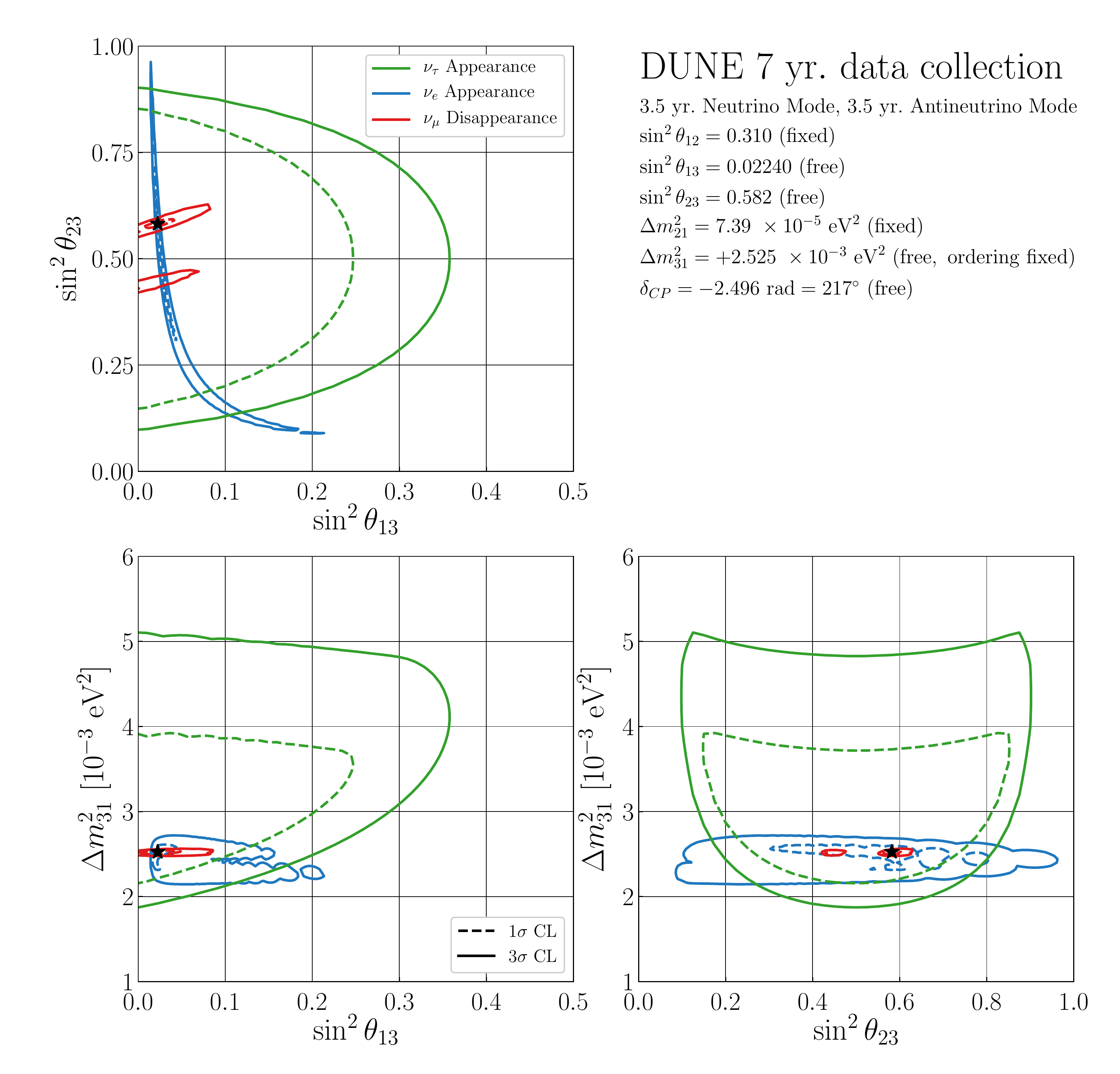}
\caption{Measurement potential of seven years of data collection at DUNE, assuming $3.5$ years each in neutrino and antineutrino modes. We show measurements assuming separate analyses of the three oscillation channels: $\nu_e$ Appearance (blue), $\nu_\tau$ Appearance (green), and $\nu_\mu$ Disappearance (red). We show $1\sigma$ and $3\sigma$ CL regions of these measurements. We fix $\sin^2\theta_{12}$ and $\Delta m_{21}^2$ to their best-fit values, assume the mass-ordering is known to be normal, and marginalize over $\delta_{CP}$. Results assuming $3$ years each in neutrino/antineutrino modes and $1$ year in high-energy mode are qualitatively similar.}
\label{fig:three-flavors}
\end{figure}

Within the three-massive-neutrinos paradigm, $\nu_{\tau}$-appearance can also be used to constrain some of the other oscillation parameters. Fig.~\ref{fig:three-flavors} depicts the allowed regions of parameters space obtained after $3.5+3.5$ years of DUNE data taking, including information from $\nu_{\tau}$-appearance (green), $\nu_{e}$-appearance (blue), and $\nu_{\mu}$-disappearance (red). Contours represent the one (dashed) and three (solid) sigma preferred regions in the $\sin^2\theta_{13}\times \sin^2\theta_{23}$, $\sin^2\theta_{13}\times \Delta m^2_{31}$, and $\sin^2\theta_{23}\times \Delta m^2_{31}$ planes. We do not depict two-dimensional projections of the parameter space that include the $\delta_{CP}$-direction since the $\nu_{\tau}$-appearance channel is virtually insensitive to $\delta_{CP}$. In all analyses, the solar parameters and the mass ordering are held fixed at their current best-fit values, while we marginalize over all other oscillation parameters not depicted. The stars indicate the best-fit points. The well-known complementarity of the $\nu_{e}$-appearance and $\nu_{\mu}$-disappearance channels is apparent, while the impact of the $\nu_{\tau}$-appearance channel is marginal when it comes to measuring oscillation parameters within the three-massive-neutrinos paradigm.
We repeated this exercise in the $3+3+1$ case and obtained results that are very similar to those of the $3.5+3.5$ case. The $3+3+1$ results are not depicted here. 

\subsection{Non-Unitary Mixing Matrix}
\label{sec:nonunitary}

One of the fundamental basic questions one can ask of neutrino oscillations is whether the lepton mixing matrix is unitary. Unitarity tests of the quark mixing matrix, for example, were a key component of the experimental and phenomenological quark flavor-physics program and stand as one of the landmarks of contemporary particle physics. Unitarity tests of the lepton mixing matrix are still rather unimpressive \cite{Qian:2013ora,Parke:2015goa} and the fact that there is very little direct information concerning $\nu_{\tau}$ appearance -- plus there is literally no information on $\nu_{\tau}$-disappearance -- contributes significantly to the current state of affairs. 

The results discussed in the last section allow one to perform a simple, mostly model-independent, unitarity test. As discussed earlier, unitarity relates the effective mixing angles that govern, for the most part, the three oscillation channels accessible to DUNE and can be translated into the following sum rule: $\sin^22\theta_{\mu\tau}+\sin^22\theta_{\mu e}-\sin^22\theta_{\mu\mu}=0$. Fig.~\ref{fig:two_param} reveals this sum rule can be tested at, roughly, the 20\% level. We performed a fit of our simulated data, assuming it is consitent with the three-massive-neutrinos paradigm, for $|U_{e3}|^2+|U_{\mu3}|^2+|U_{\tau3}|^2$, for fixed values of the solar parameters, and the mass ordering, and marginalizing over the CP-odd phase $\delta_{CP}$ and the mass-squared difference $\Delta m^2_{31}$. We find that DUNE should be able to measure, in the $3.5+3.5$ case,
\begin{equation}
|U_{e3}|^2+|U_{\mu3}|^2+|U_{\tau3}|^2=1^{+0.05}_{-0.06}~(1~\sigma)~[{\rm or}~|U_{e3}|^2+|U_{\mu3}|^2+|U_{\tau3}|^2=1^{+0.13}_{-0.17} ~(3~\sigma)],
\end{equation}
a significant improvement over the current constraints \cite{Parke:2015goa}. The precision is dominated entirely by the $\nu_{\tau}$-appearance data. If one allows for different values of $\Delta m^2_{31}$ for the different appearance channels, we estimate DUNE would measure $|U_{e3}|^2+|U_{\mu3}|^2+|U_{\tau3}|^2=1^{+0.07}_{-0.20}~(1~\sigma)$ or $1^{+0.15}_{-0.32}~(3~\sigma)$. 

In order to assess more quantitively how well the unitarity of the mixing matrix can be probed, we adopt the following parameterization for the not-necessarily-unitary lepton mixing matrix~\cite{Escrihuela:2015wra}:
\begin{equation}
U \to NU = \left(\begin{array}{c c c}\alpha_{11} & 0 & 0 \\ \alpha_{21} & \alpha_{22} & 0 \\ \alpha_{31} & \alpha_{32} & \alpha_{33} \\ \end{array}\right) U, \label{eq:NonUnitary}
\end{equation}
where $U$ is a unitary matrix. The diagonal $\alpha_{ii}$ ($i=1,2,3$) are real and the off-diagonal $\alpha_{ij} \equiv |\alpha_{ij}|e^{-\phi_{ij}}$ ($i\neq j=1,2,3$) are complex, totaling nine additional free parameters. A unitary lepton mixing matrix corresponds to $\alpha_{ij}=\delta_{ij}$. The most conservative constraints on the off-diagonal parameters are of order $|\alpha_{ij}| \lesssim 10^{-2}$, while the diagonal parameters are constrained to be $(1-\alpha_{11}) < 2.4 \times 10^{-2}$, $(1-\alpha_{22}) < 2.2 \times 10^{-2}$, $(1-\alpha_{33}) < 1.0 \times 10^{-1}$~\cite{Fernandez-Martinez:2016lgt,Escrihuela:2016ube}. In certain model-dependent situations, more stringent constraints apply~\cite{Declais:1994su,Abe:2014gda,Astier:2003gs,Astier:2001yj,Blennow:2016jkn}.

Simply stating that the lepton mixing matrix is not unitary, is not, strictly speaking, sufficient when it comes to computing the effects of nontrivial $\alpha_{ij}$ on neutrino oscillations. Here, we will restrict the discussion to the case when the $3\times 3$ mixing matrix is not unitary because there are more than three neutrino species and the new neutrino mass eigenstates are heavy enough not to be kinematically accessible in the LBNF beamline. This happens, roughly, for heavy neutrino masses heavier than the kaon mass. We also assume the new neutrino flavor eigenstates are sterile, i.e., they do participate in the electroweak interactions.  

Under these circumstances, it is convenient to express the unitary $n\times n$ lepton mixing matrix ($n=3+k$, where $k$ is the number of extra neutrino species) in block form
\begin{equation}
 U_{n\times n}=\left(\begin{array}{cc}
                V & W \\
                S & T
               \end{array}
\right),
\end{equation}
where $V\equiv NU$ is a $3\times 3$ matrix. In the mass basis, integrating out the heavy neutrino degrees of freedom (and the heavy gauge bosons),
\begin{equation}
 {\cal L}\supset\bar{\nu}_i (i\slashed{\partial}-m_i)\nu_i+\frac{4G_F}{\sqrt{2}}\bar{\ell}_{\alpha}\gamma_{\mu}P_LV_{\alpha i}\nu_{i}J^{\mu}_{\rm CC} + \frac{4G_F}{\sqrt{2}}\bar{\nu}_{i}V^{\dagger}_{i\alpha}\gamma_{\mu}P_LV_{\alpha j}\nu_jJ^{\mu}_{\rm NC}+ H.c.~,
  \label{eq:lagran_mass}
\end{equation}
where $i,j=1,2,3$, $\alpha,\beta=e,\mu,\tau$, $G_F$ is the Fermi constant, and $P_L$ is the left-chiral projection operator. $J^\mu_{\rm CC}$ and $J^\mu_{\rm NC}$ are the relevant hadronic currents for the charged-current and neutral-current weak interactions, respectively. The fact that $V$ is not unitary will impact neutrino production and lead to, for example, zero-baseline ``flavor-change,'' and modify neutrino propagation through matter. Here, we are most interested in the latter and will comment briefly on the former. As far as matter effects are concerned, the modified effective Hamiltonian used to describe neutrino flavor-evolution as a function of the baseline is, in the mass-eigenstate basis,
\begin{equation}
 H_{3\times 3}=\frac{1}{2E}{\rm Diag}\left[m_1^2,m_2^2,m_3^2\right]+U^{\dagger}N^\dagger {\rm Diag}[A_{CC}-A_{NC},-A_{NC},-A_{NC}] NU~,
\end{equation}
where  $A_{CC}$ and $A_{NC}$ are the standard charged-current and neutral-current matter potentials,  respectively.

Keeping all this in mind, we analyze our simulated data samples allowing for the hypothesis that the $N$ matrix is non-trivial. As in the previous subsection, we fix the neutrino mass ordering and the solar parameters, and marginalize over $\sin^2 \theta_{13}$, $\sin^2\theta_{23}$, $\Delta m_{31}^2$, and $\delta_{CP}$. The oscillation probability $P(\nu_\mu\to\nu_\tau)$ is most sensitive to the lowest row of $N$, the parameters $\alpha_{31}$, $\alpha_{32}$, and $\alpha_{33}$. For oscillations in vacuum and in the limit $\sin{\theta_{13}} \to 0$, there is, however, no sensitivity to the parameter $\alpha_{31}$ in the $\nu_\tau$ appearance channel. We do not, therefore, expect sensitivity to $|\alpha_{31}|$ competitive with the existing upper bound of a  few $\times 10^{-2}$, and  fix $\alpha_{31}$ to zero in our analysis. We also allow the phase of $\alpha_{32}$ to vary.

\begin{figure}[ht]
\centering
\includegraphics[width=0.65\linewidth]{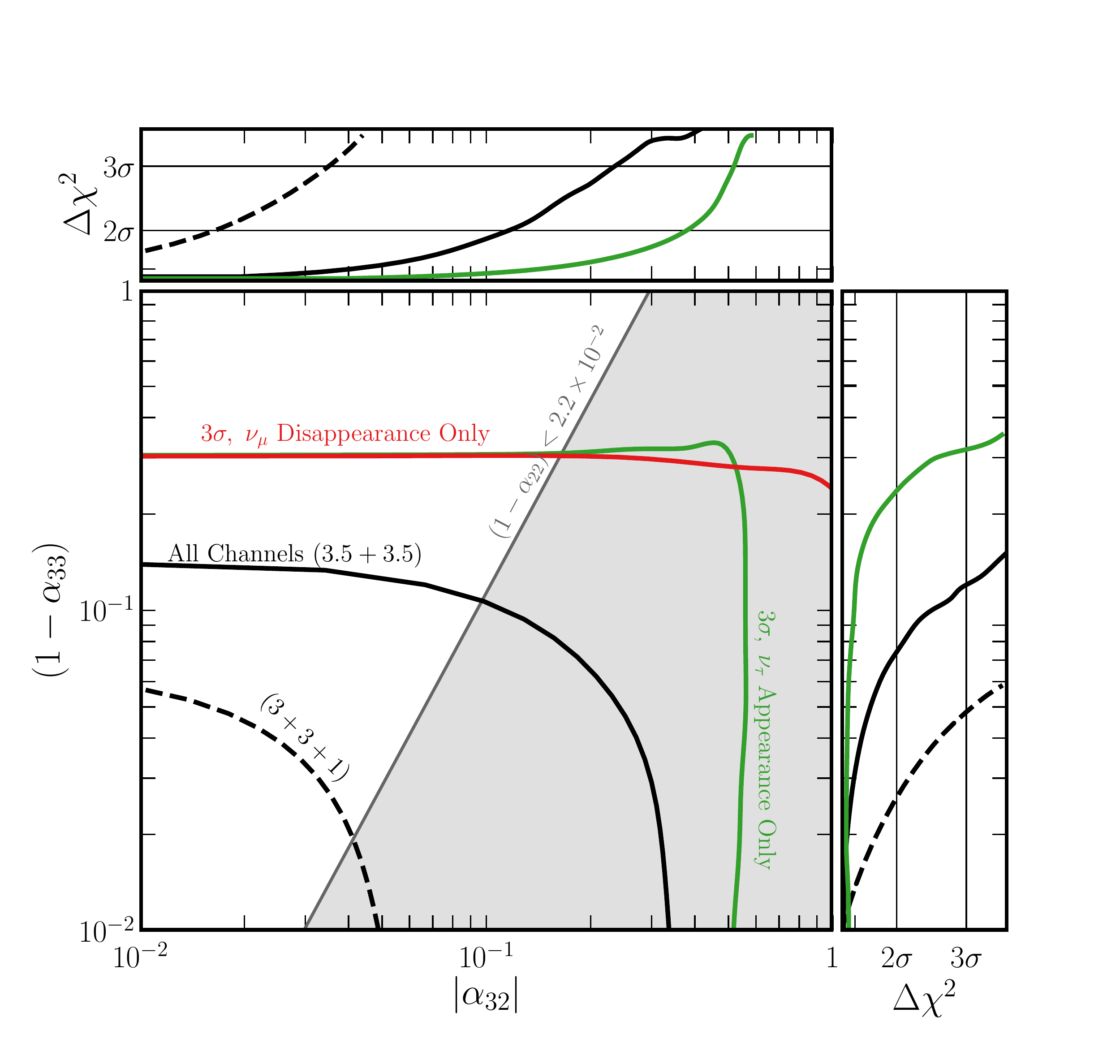}
\caption{Expected DUNE sensitivity to the non-unitarity parameters $|\alpha_{32}|$ and $(1-\alpha_{33})$. We show the $3\sigma$ CL sensitivity using the $\nu_\mu$ disappearance channel only in red and the $\nu_\tau$ appearance channel only in green. The black curves show sensitivity assuming a joint analysis of all channels, assuming $3.5$ years each in neutrino and antineutrino modes (solid line) or $3$ years each in neutrino and antineutrino modes, as well as $1$ year of high-energy mode (dashed line). One-dimensional $\Delta \chi^2$ projections for each parameter are shown above ($|\alpha_{32}|$) and to the right ($1 - \alpha_{33}$). The grey region is ruled out by constraints on the parameter $(1 - \alpha_{22}$).}
\label{fig:NonUnitary}
\end{figure}

Fig.~\ref{fig:NonUnitary} depicts (green contour) the expected sensitivity of the $\nu_{\tau}$-appearance channel at DUNE in the $|\alpha_{32}| \times(1-\alpha_{33})$-plane, in the $3.5+3.5$ case. It also depicts the sensitivity of the $\nu_{\mu}$ disappearance channel (red contour), and the reach of a combined analysis of all channels (solid black contour). We also show expected sensitivity of the combined analysis assuming $3+3+1$ years of data collection (dashed black contour). In the parameterization of non-unitarity using $\alpha_{ij}$, the parameters are not completely independent: the off-diagonal elements $|\alpha_{ij}|$ must be less than the product $2\sqrt{(1-\alpha_{ii})(1-\alpha_{jj})}$. The existing constraint $(1-\alpha_{22}) < 2.2 \times 10^{-2}$ therefore excludes combinations of $|\alpha_{32}|$ and $(1-\alpha_{33})$ in the grey shaded region of Fig.~\ref{fig:NonUnitary}. The expected limit on $|\alpha_{32}|$ is considerably weaker than existing limits, while that on $(1-\alpha_{33})$ is comparable to the $\mathcal{O}(10^{-1})$ existing limit, even after one includes information from $\alpha_{22}$. 

For higher neutrino energies, the sensitivity to the unitarity-violating parameters is expected to increase. This is due to the increased importance of matter effects for higher neutrino energies and the fact that the matter potential depends non-trivially on the unitarity-violating parameters. Fig.~\ref{fig:NonUnitary} depicts (dashed black contour) the expected combined sensitivity  in the $\alpha_{32}\times(1-\alpha_{33})$-plane, in the $3+3+1$ case. It is easy to see that in this case the sensitivity is significantly stronger than the one in the $3.5+3.5$ case. Fig.~\ref{fig:NonUnitary} reveals that, in the $3+3+1$ running scheme, we are sensitive to $(1-\alpha_{33}) > 6 \times 10^{-2}$ at the $3\sigma$ confidence level.

As mentioned earlier, a non-unitary lepton mixing matrix, assuming it arises as hypothesized here, leads to non-trivial flavor-change in the limit where the baseline vanishes. This is a consequence of the fact that, for example, the neutrino produced in $\pi^+\to\mu^++\nu$ is not orthogonal to the one that participates in $\nu+n\to p+e^-$. As far as $\nu_{\tau}$ appearance is concerned, the search for $\nu_{\tau}$ candidates in the DUNE near detector complex would be sensitive to $|\alpha_{33}|^2|\alpha_{32}|^2$. The large neutrino flux, combined with the fact that there are no beam-related backgrounds for $\nu_{\tau}$ appearance in the DUNE near detector site, should translate into exquisite sensitivity to $|\alpha_{33}|^2 |\alpha_{32}|^2$. The pursuit of this and other near-detector-related question, is beyond the aspirations of this manuscript. See Refs.~\cite{Hernandez-Garcia:2017pwx,Miranda:2018yym,Dutta:2019hmb} for discussions of non-unitarity effects at near detectors not involving $\nu_\tau$ appearance.

\setcounter{footnote}{0}
\subsection{The Three Plus One Neutrino Hypothesis}
\label{sec:sterile}

In the previous subsection, we discussed the sensitivity of $\nu_{\tau}$-appearance to the hypothesis that there are heavy neutrino states. If, instead, the new neutrino states are light -- much lighter than the pion -- they cannot be integrated out but must be included among the kinematically accessible neutrino mass eigenstates. The so-called 3+1 scenario has been the subject of immense phenomenological scrutiny, and remains a possible solution, albeit under significant stress, to the so-called short-baseline anomalies. In this subsection, we discuss how well the $\nu_{\tau}$-appearance data sample to be collected at DUNE can constrain the 3+1 neutrino hypothesis, and how its information complements that which can be obtained via the $\nu_e$-appearance or the $\nu_{\mu}$-disappearance channels.

In more detail, we add a fourth mass-eigenstate, with mass $m_4$, to the three existing neutrino mass eigenstates. We further assume that the new interaction eigenstate is sterile, i.e., does not participate in charged-current or neutral-current weak interactions. In this case, the mixing matrix is $4\times 4$, and, in vacuum and in the limit $\Delta m_{12}^2 \to 0$, 
\begin{eqnarray}
P(\nu_\mu \to \nu_\tau) & = & 4|U_{\mu 3}|^2 |U_{\tau 3}|^2 \sin^2{\left(\frac{\Delta m_{13}^2 L}{4E_\nu}\right)} + 4|U_{\mu 4}|^2 |U_{\tau 4}|^2 \sin^2{\left(\frac{\Delta m_{14}^2 L}{4E_\nu}\right)} \nonumber \\
& + & 4|U_{\mu 3} U_{\tau 3} U_{\mu 4} U_{\tau 4}| \sin{\left(\frac{\Delta m_{13}^2 L}{4E_\nu}\right)} \sin{\left(\frac{\Delta m_{14}^2 L}{4E_\nu}\right)} \cos{\left(\frac{\Delta m_{34}^2 L}{4E_\nu} + \delta_{24}\right)},  \label{eq:3+1}
\end{eqnarray}
where $\delta_{24}$ is the new $CP$-violating phase associated with $U_{\mu 4}$ (we use the parameterization from Ref.~\cite{Berryman:2015nua}). The first term in Eq.~(\ref{eq:3+1}) is the standard ``atmospheric'' term, Eq.~(\ref{eq:pmutau}), while the extra contributions are proportional to $|U_{\mu4}U_{\tau4}| = \sin{\theta_{24}}\sin{\theta_{34}}\cos^2{\theta_{14}}\cos{\theta_{24}} \simeq \sin{\theta_{24}} \sin{\theta_{34}}$ in the limit where  all ``new'' mixing angles are small.

We simulate data as in the previous subsection and analyze it assuming the 3+1 neutrino hypothesis. Here, however, we include priors on $\Delta m_{31}^2$ and $|U_{\mu 3}|^2$ from the T2K Experiment~\cite{Abe:2018wpn}, $|U_{e3}|^2$ from the Daya Bay experiment~\cite{Adey:2018zwh}, and the solar parameters ($|U_{e2}|^2$ and $\Delta m_{21}^2$) from the solar experiments and KamLAND~\cite{Esteban:2018azc}.\footnote{The reason for using a subset of existing bounds is one of convenience. These are the same priors we use in the next subsection, when we are interested in the subset of experiments that are not expected to be signifciantly impacted by non-standard neutrino--matter neutral-current interactions.} Note that the priors are imposed on the magnitudes of some elements of the mixing matrix, not on the values of the mixing angles. For a more detailed discussion on how these bounds are imposed, see \cite{Berryman:2015nua}. The reason for the priors is as follows. We want to gauge the impact of $\nu_{\tau}$-appearance at DUNE without the benefit of information from the other oscillation channels. On the other hand, the $\nu_{\tau}$-appearance channel, as we argued earlier, provides only very limited information regarding the dominant oscillation frequency and, for the most part, is unable to distinguish one from several oscillation frequencies. The priors impose the constraint that there are two known oscillation frequencies -- proportional to $\Delta m^2_{31}$ and $\Delta m^2_{21}$ and that both $\nu_{\mu}$ and $\nu_e$ ``see'' these two frequencies. Both are well established facts. This way, we infer what we believe is a more meaningful estimate of the sensitivity.

\begin{figure}[ht]
\center
\includegraphics[width=0.55\linewidth]{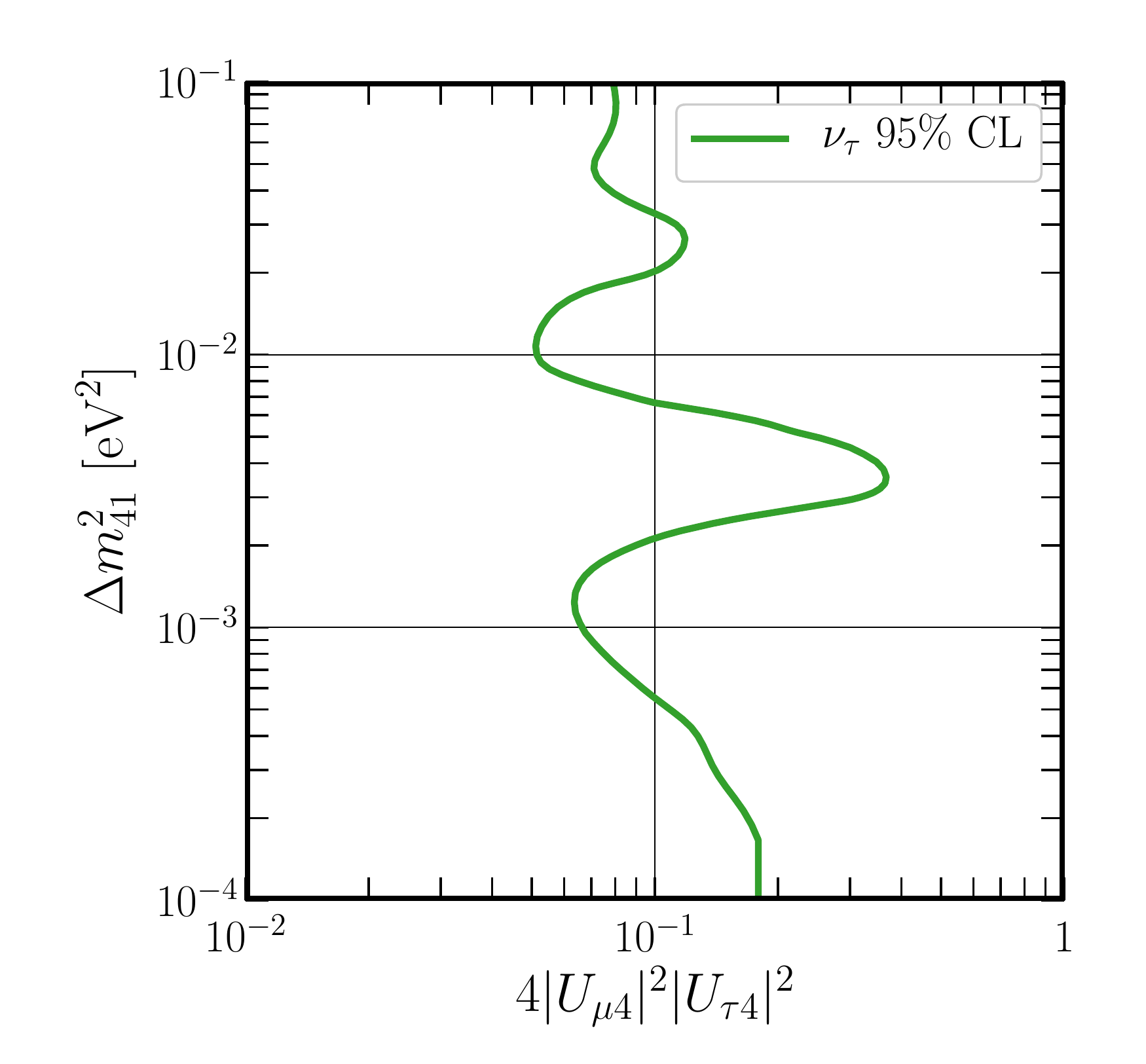}
\caption{Expected 95\% CL sensitivity to the combination of mixing matrix elements $4|U_{\mu 4}|^2 |U_{\tau 4}|^2$ vs. $\Delta m_{41}^2$ using only the $\nu_\tau$ appearance channel, assuming $3.5$ years of data collection each in neutrino and antineutrino modes. This analysis includes priors on the solar parameters ($|U_{e2}|^2$ and $\Delta m_{21}^2$) from the solar experiments and KamLAND~\cite{Esteban:2018azc}, priors on $\Delta m_{31}^2$ and $|U_{\mu 3}|^2$ from the T2K Experiment~\cite{Abe:2018wpn}, and $|U_{e3}|^2$ from the Daya Bay experiment~\cite{Adey:2018zwh}, as discussed in the text. It also fixes the mass ordering to be normal. All other parameters, including the additional mixing angles and $CP$-violating phases are marginalized over in our analysis.}
\label{fig:SimpleSterile}
\end{figure}

Fig.~\ref{fig:SimpleSterile} (left) depicts the sensitivity, in the $3.5+3.5$ case, of DUNE, in the $4|U_{\mu4}|^2|U_{\tau4}|^2\times \Delta m^2_{41}$-plane. We fixed the ``123'' neutrino mass ordering -- normal -- and the solar parameters, and marginalize over all other oscillation parameters. We restrict the parameter space to $m_4^2>m_1^2$ and restrict our analysis to values of $\Delta m^2_{41}$ less than 0.1~eV$^2$. For larger values of $\Delta m^2_{41}$, we expect $\nu_{\tau}$-appearance searches in the near detector -- beyond the scope of this manuscript -- to play a decisive role. For $\Delta m^2_{41}\lesssim 10^{-4}$~eV$^2$ (or lower),  the sensitivity of DUNE does not depend on $\Delta m^2_{41}$.

Current data constrain $|U_{\tau4}|$ only rather poorly and mostly indirectly. There are, however, significant constraints on $|U_{\mu 4}|$, mostly from $\nu_{\mu}$ disappearance searches. The MINOS and MINOS+ collaborations~\cite{Adamson:2017uda} constrain $|U_{\mu4}|^2\lesssim 10^{-2}$ for $\Delta m^2_{41}\gtrsim 10^{-3}$~eV$^2$ and $|U_{\mu4}|^2\lesssim 0.5$ for $\Delta m^2_{41}\ll10^{-3}$~eV$^2$. Significantly better sensitivity is expected from the disappearance channel at DUNE \cite{Berryman:2015nua,Gandhi:2015xza,Dutta:2016glq,Gupta:2018qsv}. Matter effects also lead to some sensitivity to $|U_{\tau 4}|$ in the $\nu_{\mu}$-disappearance and $\nu_e$-appearance channels~\cite{Abe:2014gda,Aartsen:2017bap,Blennow:2018hto}. Inspecting Fig.~\ref{fig:SimpleSterile}, one would be tempted to believe, naively, that the current (and future, assuming no discovery) bounds on $|U_{\mu4}|^2$ would render the bounds on $|U_{\tau 4}|^2$ from $\nu_{\tau}$-appearance at DUNE trivial,  especially for $\Delta m^2_{41}\gtrsim 10^{-3}$~eV$^2$. This is not correct. Even in the limit $|U_{\mu4}|^2\to 0$, the $\nu_{\tau}$-appearance data, assuming it is consistent with the three-massive-neutrinos paradigm, can rule out large $|U_{\tau4}|^2$ via the (unitarity) sum rule $|U_{\tau4}|^2\le 1-|U_{\tau3}|^2$. The lower bound on $|U_{\tau3}|^2$ translates into a robust bound on $|U_{\tau4}|^2$.

\begin{figure}[ht]
\centerline
{\includegraphics[width=1\textwidth]{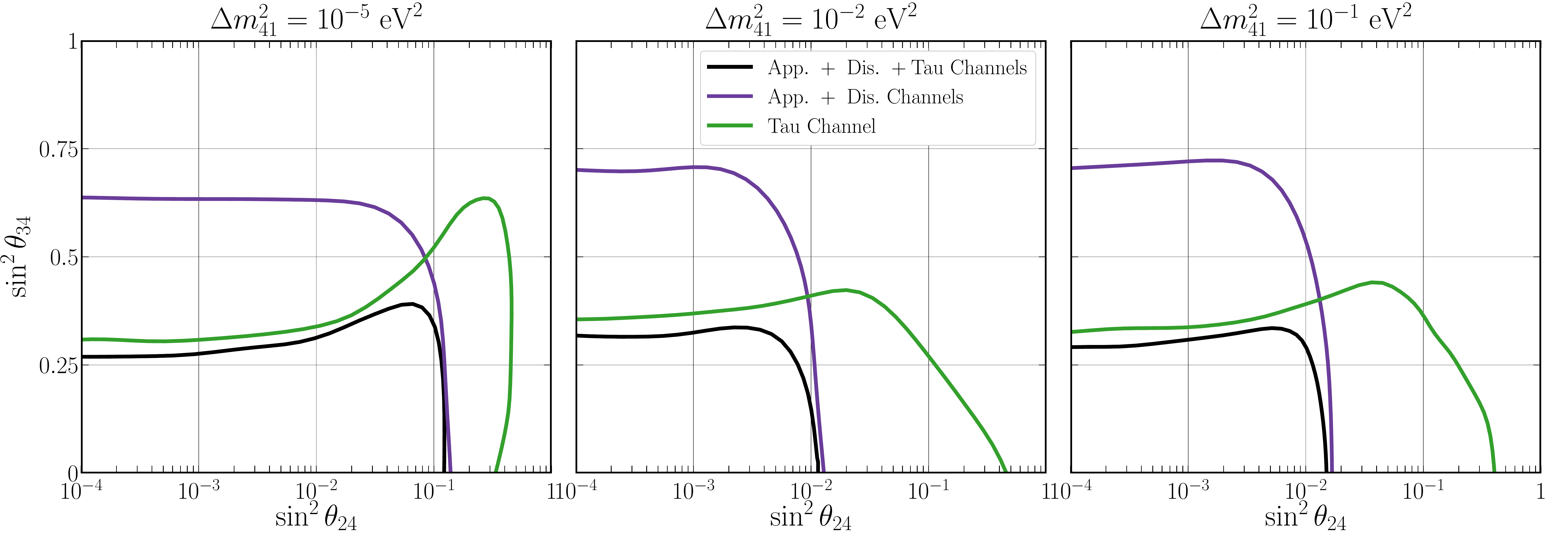}}
\caption{Expected sensitivity to the new mixing angles $\sin^2\theta_{24}$ and $\sin^2\theta_{34}$ for three different fixed values of the new mass-squared difference $\Delta m_{41}^2 = 10^{-5}$ eV$^2$ (left), $10^{-2}$ eV$^2$ (center), and $10^{-1}$ eV$^2$ (right). We compare sensitivity using $\nu_\tau$ appearance data only (green) with a combined $\nu_e$ appearance $\nu_\mu$ disappearance analysis (purple), as well as a joint analysis (black). Solar parameters are fixed to their best-fit values, the mass-ordering is fixed to normal, and all other unseen parameters have been marginalized over.}
\label{fig:24_34}
\end{figure}

Figure~\ref{fig:24_34} depicts, in the $3.5+3.5$ case, the sensitivity of DUNE to the fourth light-neutrino hypothesis, this time in the $\sin^2\theta_{24}\times \sin^2\theta_{34}$ plane, for fixed values of the new mass-squared difference. In the left panel, $\Delta m^2_{41}$ is low enough that oscillations driven by the new mass-squared difference have not yet developed at the DUNE far detector. In the middle panel, $\Delta m_{41}^2$ is close to $\Delta m_{31}^2$, meaning oscillations are relevant at DUNE. In the right panel, the new oscillations have averaged out at the far detector. Information from the $\nu_{\tau}$-appearance channel (green) complements that from the combined $\nu_{\mu}$-disappearance and $\nu_e$-appearance channels (purple). The sensitivity of combined analyses is depicted in black. In the case of small $\sin^2\theta_{24}$, constraints from $\nu_{\tau}$-appearance on $\sin^2\theta_{34}$ always surpass those from the other oscillation channels, for small, intermediate, or large values of $\Delta m^2_{41}$. We note that the results depicted in Figure~\ref{fig:24_34} are less sensitive to the priors on $\Delta m^2_{31}$, $|U_{e3}|^2$, and $|U_{\mu3}|^2$ than those depicted in Fig.~\ref{fig:SimpleSterile}.

We repeated this exercise in the $3+3+1$ case and obtained results -- not presented here for the sake of conciseness -- that are very similar to those of the $3.5+3.5$ case. 

\begin{figure}[ht]
\center
\includegraphics[width=0.55\linewidth]{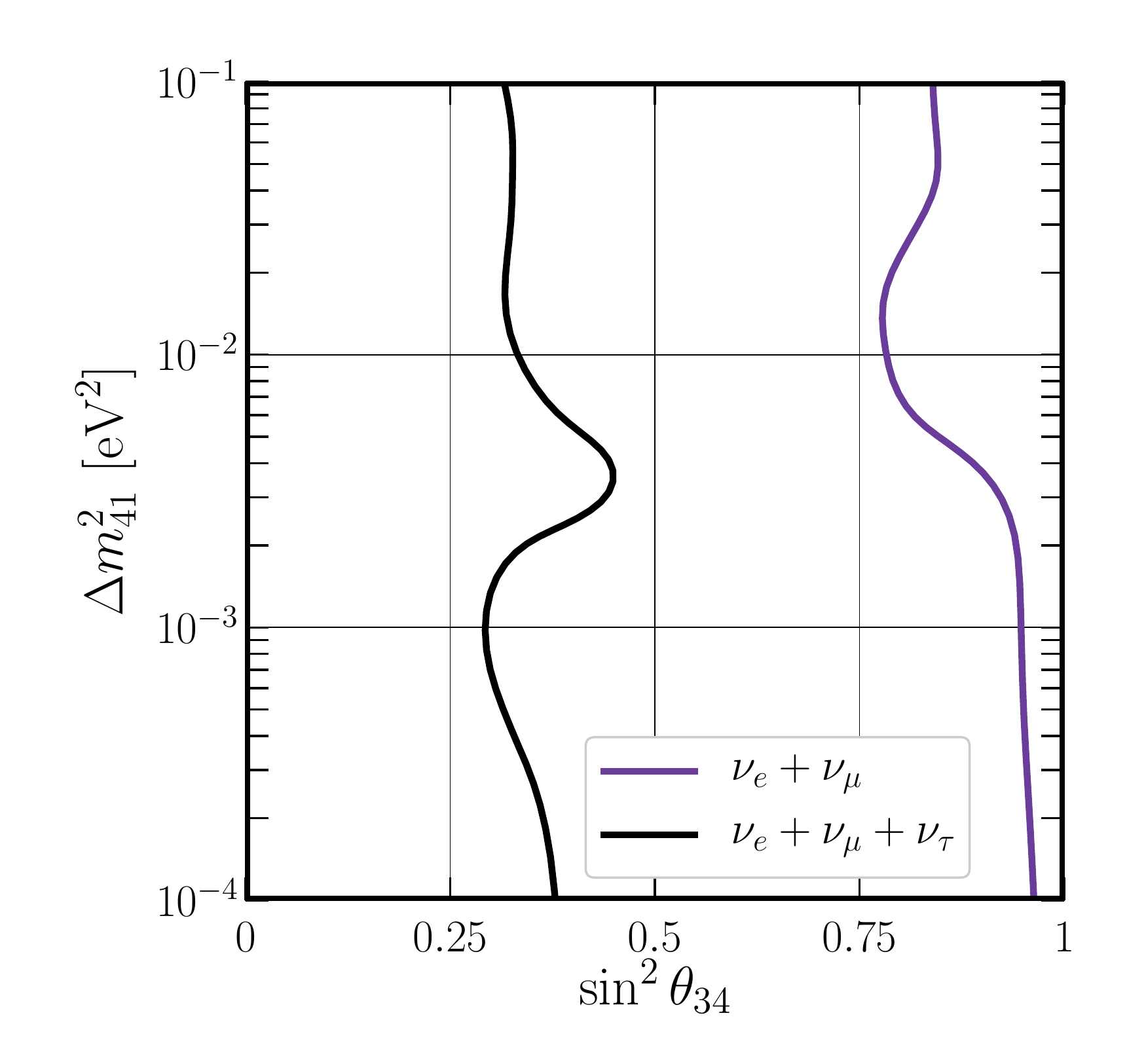}
\caption{Expected 95\% CL sensitivity in the  $\sin^2\theta_{34}\times\Delta m^2_{41}$-plane,  analyzing only $\nu_e$-appearance and $\nu_\mu$-disappearance data (purple) or further including the $\nu_\tau$-appearance channel (black). Note the linear scaling on $\sin^2\theta_{34}$. We include external priors on the solar parameters ($|U_{e2}|^2$ and $\Delta m_{21}^2$) from the solar experiments and KamLAND~\cite{Esteban:2018azc} and fix the mass ordering to be normal. All other parameters, including the additional mixing angles and $CP$-violating phases are marginalized over.}
\label{fig:Sterile_new}
\end{figure}

In order to further illustrate the impact of the $\nu_{\tau}$-appearance data on the DUNE's capability to test the 3+1 hypothesis, we addressed a slightly different question: how does the sensitivity of DUNE change once $\nu_{\tau}$-appearance data are added to the more traditional $\nu_{e}$-appearance and $\nu_{\mu}$-disapparance samples? We concentrate our discussion on $|U_{\tau4}|^2\propto\sin^2\theta_{34}$. Fig.~\ref{fig:Sterile_new} depicts, in the $3.5+3.5$ case,  the sensitivy of DUNE in the $\sin^2\theta_{34}\times \Delta m^2_{41}$-plane when analyzing only $\nu_e$-appearance and $\nu_\mu$-disappearance (purple) or further including information from the $\nu_\tau$-appearance data (black). We draw attention to our use of a linear scale for $\sin^2\theta_{34}$. Here, unlike the analysis that led to Figs.~\ref{fig:SimpleSterile}, and \ref{fig:24_34},  we only make use of external priors on $|U_{e2}|^2$ and $\Delta m_{21}^2$~\cite{Esteban:2018azc}. The impact of the $\nu_{\tau}$-appearance data is nontrivial. 

\subsection{Non-standard Neutrino Interactions}
\label{sec:NSI}

New interactions between neutrinos and ordinary matter can also modify neutrino propagation in a way that is testable at long-baseline experiments. In this subsection, we consider additional neutral-current-like interactions between neutrinos and matter (electrons, up quarks, and down quarks) -- non-standard neutrino interactions (NSI) -- that modify the matter potential for neutrino oscillations. The modified matter potential is traditionally written, in the flavor basis, as
\begin{equation}\label{eq:NSI}
 A\left(\begin{array}{c c c} 1+\epsilon_{ee} & \epsilon_{e\mu} & \epsilon_{e\tau} \\ \epsilon_{e\mu}^* & \epsilon_{\mu\mu} & \epsilon_{\mu\tau} \\ \epsilon_{e\tau}^* & \epsilon_{\mu\tau}^* & \epsilon_{\tau\tau}\end{array}\right),
\end{equation}
where $\epsilon_{\alpha\beta}$ are reweighted quantities, dependent on the strength of new interactions between neutrinos of flavor $\alpha$ and $\beta$ with constituent elements of the Earth along the path of propagation (see, e.g., Refs.~\cite{Friedland:2004pp,Friedland:2004ah,Friedland:2005vy,Yasuda:2010hw,GonzalezGarcia:2011my,Choubey:2014iia,Friedland:2006pi,Ohlsson:2012kf,Kikuchi:2008vq,deGouvea:2015ndi,Farzan:2017xzy}), and $A = \sqrt{2} G_F n_e$, with $n_e$ being the number density of electrons along the path of propagation. For propagation close to the surface of the Earth, $A \simeq 6\times 10^{-4}$ eV$^2$/GeV. This is to be compared to the different oscillation frequencies, proportional to $\Delta m_{21}^2/E$ and $\Delta m_{31}^2/E$.

The NSI parameters are complex, while the flavor-propagation Hamiltonian is Hermitian, leading to a total of nine free parameters. Neutrino oscillations are insensitive to an overall shift to all the eigenvalues of the propagation Hamiltonian, so we may subtract off any NSI parameter times the identity -- in practice this is typically $\epsilon_{\mu\mu}$, as external constraints on this parameter are quite strong~\cite{Ohlsson:2012kf}. Hence, limits from oscillations are usually derived for $\epsilon_{ee} - \epsilon_{\mu\mu}$ and $\epsilon_{\tau\tau} - \epsilon_{\mu\mu}$. In practice, for the sake of DUNE, the existing strong limits on $\epsilon_{\mu\mu}$ imply that there is effectively no difference between the limits on $\epsilon_{\alpha\alpha} - \epsilon_{\mu\mu}$ and $\epsilon_{\alpha\alpha}$ ($\alpha=e,\tau$), so we will make no distinction between the two in what follows. For oscillation-based probes\footnote{The effective operators that lead to Eq.~(\ref{eq:NSI}) are not manifestly gauge-invariant. If one assumes $SU(2)_L\times U(1)_Y$ gauge invariance, lepton-flavor-violating processes and their non-observation normally provide very strong constraints on the NSI parameters~\cite{Davidson:2003ha,Abdallah:2003np,Ribeiro:2007ud}.}, the existing constraints on NSI parameters are at the level of $\mathcal{O}(10^{-1})-\mathcal{O}(10)$, the weakest being on the $\epsilon_{\alpha\tau}$ parameters. 

The sensitivity of an experiment to NSI parameters is dependent on which oscillation probability is being measured and, in general, the channel $P(\nu_\alpha \to \nu_\beta)$ is most strongly dependent on the $\epsilon_{\alpha\beta}$ parameters ($\alpha=e,\mu,\tau$). Since the $\nu_e$ appearance measurements at DUNE are quite sensitive to matter effects -- the $1$ in the $ee$-component of Eq.~(\ref{eq:NSI}) -- we expect DUNE to be sensitive to $\mathcal{O}(0.1)$ values of the NSI parameters. Refs.~\cite{deGouvea:2015ndi,Coloma:2015kiu} explored the ability of the $\nu_e$ appearance and $\nu_\mu$ disappearance channels to probe these NSI parameters in great detail.

Writing the oscillation probabilities in matter with nonzero $\epsilon_{\alpha\beta}$ is a nontrivial exercise, and many perturbative approaches have been developed to express the probabilities analytically~\cite{Kikuchi:2008vq,Martinez-Soler:2018lcy,Chaves:2018sih}. In general, for the baselines and energies of interest at DUNE, the probability $P(\nu_\mu \to \nu_\tau)$ depends predominantly on the parameters $\epsilon_{\mu\tau}$ and $\epsilon_{\tau\tau}$. Moreover, the oscillation probability is more sensitive to changes to the real part of $\epsilon_{\mu\tau}$ than the imaginary part, so we expect sensitivity to be weakest when $\phi_{\mu\tau}$, the phase associated with $\epsilon_{\mu\tau}$ is $\pi/2$ or $3\pi/2$. In the $\nu_\tau$ appearance channel, we expect sensitivity to $\epsilon_{e\mu}$ and $\epsilon_{e\tau}$ only if the parameters are large ($\gtrsim 1$). Additionally, in the $\epsilon_{ee}\gg 1$ limit, $\nu_e$ becomes an eigenstate of the flavor-evolution Hamiltonian, and all sensitivity to $\epsilon_{e\mu}$ and $\epsilon_{e\tau}$ will vanish in the $\nu_\tau$ appearance channel. For this reason, we set $\epsilon_{ee} = 0$ going forward.

\begin{figure}[ht]
\centerline{
\includegraphics[width=1\textwidth]{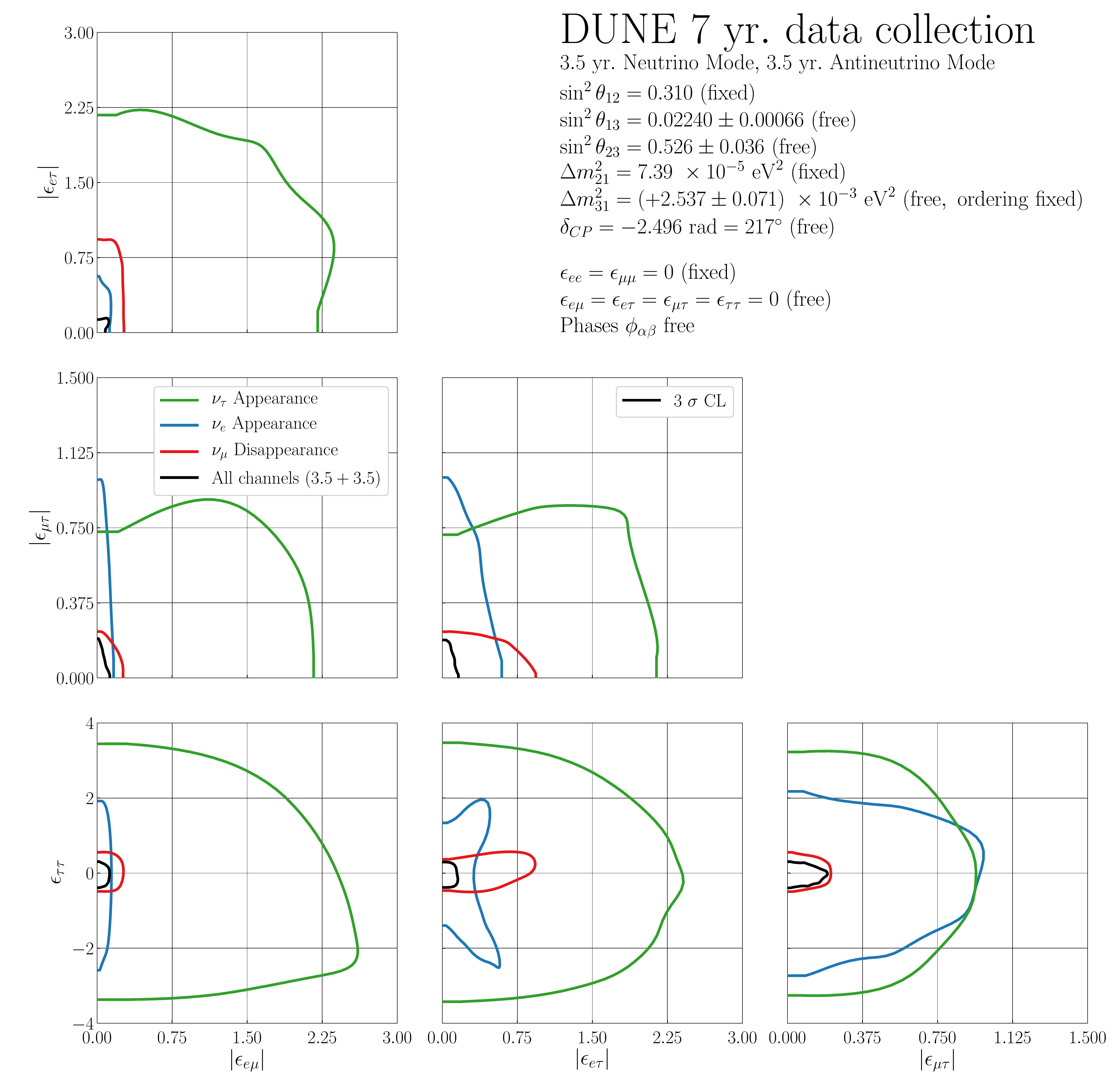}}
\caption{Expected sensitivity using $\nu_\tau$ appearance data (green), $\nu_e$ appearance data (blue), $\nu_\mu$ disappearance data (red), and a combined analysis of all three (black) at DUNE assuming 3.5 years each of neutrino and antineutrino modes. As shown in the figure and explained in the text, we have included priors on $\sin^2\theta_{13}$, $\sin^2\theta_{23}$, and $\Delta m_{31}^2$, have fixed $\sin^2\theta_{12}$ and $\Delta m_{21}^2$ to their best-fit values, and assumed the mass ordering is known to be normal. The NSI phases have also been marginalized over.}
\label{fig:NSISensitivity}
\end{figure}

Fig.~\ref{fig:NSISensitivity} depicts the expected $3\sigma$ sensitivity of the DUNE $\nu_\tau$ appearance channel (in green) to the NSI parameters $|\epsilon_{e\mu}|$, $|\epsilon_{e\tau}|$, $|\epsilon_{\mu\tau}|$, and $\epsilon_{\tau\tau}$ in the $3.5+3.5$ case. Degeneracies between NSI parameters and the three-massive-neutrinos-paradigm parameters have been extensively studied in the literature~\cite{Friedland:2004ah,Friedland:2005vy,Friedland:2006pi,Kopp:2010qt,GonzalezGarcia:2011my,Coloma:2011rq,deGouvea:2015ndi,Coloma:2015kiu,Liao:2016hsa,Bakhti:2016prn,Coloma:2016gei,deGouvea:2016pom,Blennow:2016etl,Blennow:2016jkn,Fukasawa:2016lew,Ghosh:2017ged}. To reduce the effect of such degeneracies on this analysis, we have included as priors information from existing neutrino oscillation searches that are generally insensitive to NSI. As in previous sections, we simply fix the solar neutrino parameters to their best-fit points from NuFit and assume the mass ordering is normal. Here, however, we also include a prior on $\sin^2\theta_{13} = 0.02240 \pm 0.00066$, since the results of the Daya Bay experiment are effectively insensitive to matter effects~\cite{Adey:2018zwh}. We also include priors from the T2K experiment on the measurements of $\sin^2\theta_{23} = 0.526 \pm 0.036$ and $\Delta m_{31}^2 = (2.537 \pm 0.071) \times 10^{-3}$ eV$^2$~\cite{Abe:2018wpn}. In our analysis, we keep the phases $\phi_{\alpha\beta}$ of the off-diagonal NSI parameters free, as well as $\sin^2\theta_{13}$, $\sin^2\theta_{23}$, $\delta_{CP}$, $\Delta m_{31}^2$, along with  the parameters depicted in Fig.~\ref{fig:NSISensitivity}.

For comparison, Fig.~\ref{fig:NSISensitivity} also depicts the sensitivity of the $\nu_e$ appearance channel (blue) or the $\nu_\mu$ disappearance channel (red) to the NSI parameters, subject to the same priors as in the $\nu_\tau$ channel. In black, we display the results of a combined analysis using all three channels. We see that, generically, these channels are more powerful than the $\nu_\tau$ channel, however sensitivities of individual channels can be comparable (for instance, $|\epsilon_{\mu\tau}|$ vs. $\epsilon_{\tau\tau}$, comparing $\nu_e$ and $\nu_\tau$ appearance channels). As with other probes, we stress the importance of performing independent analyses of the diferent channels as a way of cross-checking results. 

We also repeated this analysis assuming $3$ years of neutrino beam, $3$ years of antineutrino beam, and $1$ year of the high-energy, $\nu_\tau$ optimized beam, and found that the sensitivity was comparable to what is displayed in Fig.~\ref{fig:NSISensitivity}. The only noticeable difference is the $3+3+1$ analysis is slightly more sensitive to $|\epsilon_{\mu\tau}|$ than the $3.5+3.5$ one. This is due to the large number of events in the $\nu_\mu$ disappearance channel at large neutrino energy, where effects of $|\epsilon_{\mu\tau}|$ are more pronounced. We choose not to display the results of this analysis due to their similarity to those depicted in Fig.~\ref{fig:NSISensitivity}.

\setcounter{equation}{0}
\section{Discussion \& Conclusions}
\label{sec:conclusion}

As of today, only a handful of $\nu_{\tau}$ events have been directly observed via the standard model charged-current weak interactions. The LBNF-DUNE experimental setup is expected to collect and isolate a  $\nu_{\tau}$-enriched sample in the DUNE far detector that dwarfs all existing data samples in both its size and purity. Here we explored the future physics sensitivity of seven years of $\nu_{\tau}$-appearance searches in the DUNE far detector. 

The collection of a $\nu_{\tau}$-enriched sample is a nontrivial task. The high $\tau$-production threshold, a consequence of the heavy $\tau$-lepton mass,  implies relatively very small statistics, and the fact that the $\tau$ leptons decay promptly and only semi-visibly implies that reconstructing the $\nu_{\tau}$ energy is difficult and that the rejection of neutral-current backgrounds is challenging. These challenges are reflected in the results presented here. The sensitivity of the $\nu_{\tau}$-appearance channel, whenever comparisons are meaningful, is markedly inferior to that of the $\nu_{e}$-appearance or the $\nu_{\mu}$-disappearance ones as one can readily confirm in Figs.~\ref{fig:three-flavors} or \ref{fig:NSISensitivity}. On the other hand, we emphasize the complementarity of the different oscillation channels. Different hypotheses about the physics responsible for neutrino oscillations, including the very successful three-massive-neutrinos paradigm, imply different correlations among oscillation channels and, in many cases, the $\nu_{\tau}$-appearance channel provides information that cannot be accessed in other ways. 

\begin{figure}[ht]
\center
\includegraphics[width=0.65\textwidth]{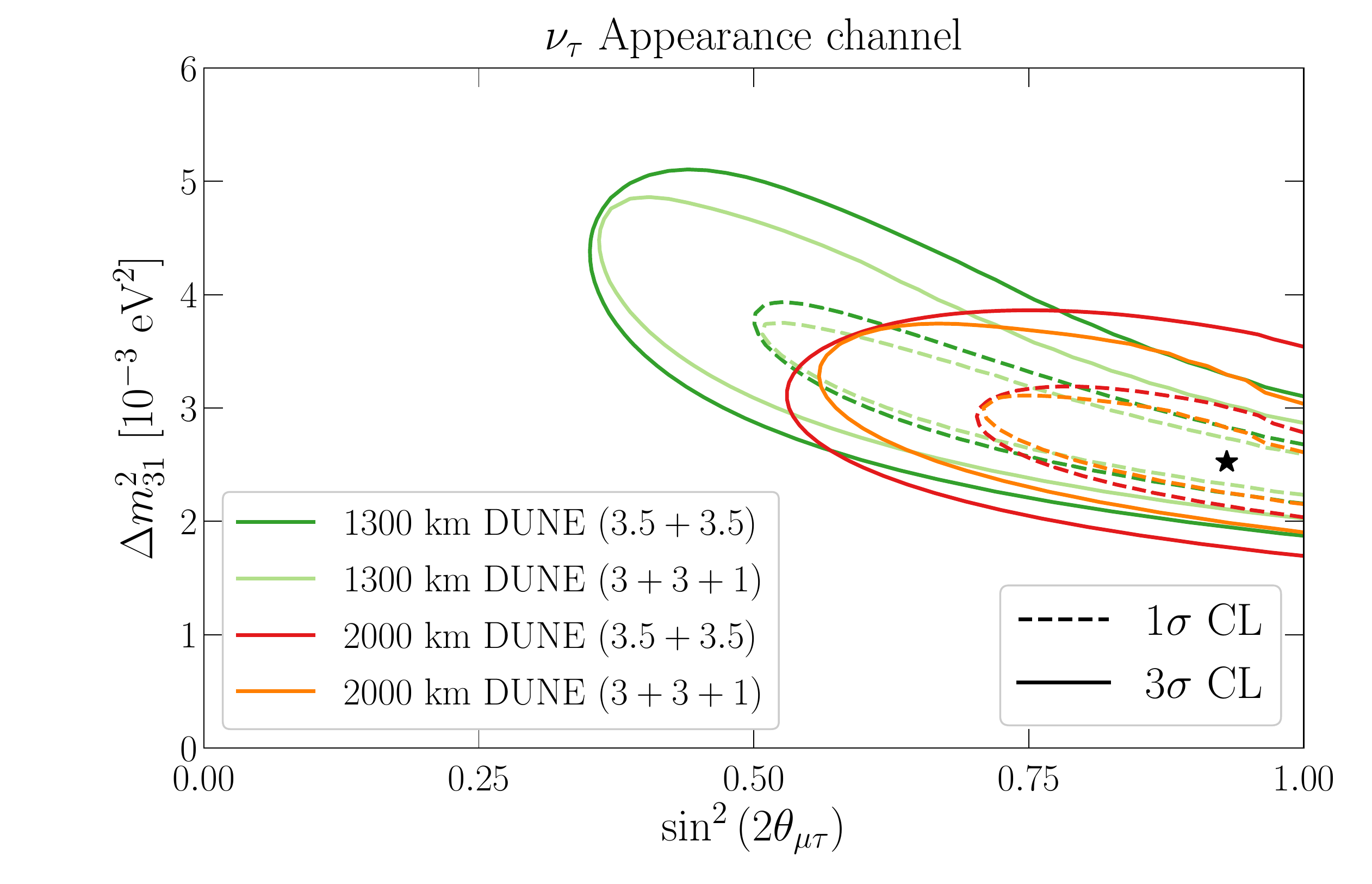}
\caption{Green lines: identical to the center panel of Fig.~\ref{fig:two_param}. The red and orange lines assume the DUNE detectors are a distance of $2000$~km from Fermilab for a $3.5 + 3.5$ and $3 + 3 + 1$ data collection scenario, respectively.}
\label{fig:2000km}
\end{figure}

Another interesting side effect of the large $\tau$-production threshold is that, in the case of the LBNF-DUNE setup, all $\nu_{\tau}$-appearance events are above the first oscillation maximum, see Fig.~\ref{fig:osci_prob}. This implies that one is not able to explore the oscillatory behavior of the oscillation probability, a fact that translates into the strong correlation between $\Delta m^2_{31}$ and $\sin^22\theta_{\mu\tau}$  in Fig.~\ref{fig:two_param}. This fact also partially explains why, in the case of the three-massive-neutrinos paradigm, data taking in the high energy configuration does not lead to qualitatively better results. To better illustrate this point, we repeated the exercise that led to  Fig.~\ref{fig:two_param} assuming that the LBNF-DUNE baseline is $L=2000$~km. The associated measurement of $\sin^22\theta_{\mu\tau}$ and $\Delta m^2_{31}$ is depicted in Fig.~\ref{fig:2000km} for both 3.5 years of running in the neutrino mode and 3.5 years of running in the antineutrino mode, and for 3 years in both neutrino and antineutrino modes and one year in the high energy beam configuration. Even though the neutrino fluxes are smaller at 2000~km (relative to 1300~km), the ``measurement'' of the oscillation parameters is more precise for the longer baseline. Furthermore, the 3+3+1 case allows for a slightly more precise measurement than the 3.5+3.5 case. 

We explored the impact of the $\nu_{\tau}$-appearance channel on testing different hypotheses: the three-massive neutrinos paradigm, the existence of new heavy neutrino degrees of freedom (nonunitarity), the existence of new light neutrino degrees of freedom (3+1 neutrino oscillations), and the existence of new neutrino--matter neutral-current interactions (NSI). Our results are summarized in Figs.~\ref{fig:three-flavors}, \ref{fig:NonUnitary}, \ref{fig:SimpleSterile}, \ref{fig:24_34}, \ref{fig:Sterile_new}, and \ref{fig:NSISensitivity}. In many cases we also depicted the sensitivity of the $\nu_{e}$-appearance or the $\nu_{\mu}$-disappearance channels. It is always instructive to appreciate the limitations of the different oscillation channels and how they complement one another. We also compared two different running schemes for LBNF-DUNE -- the 3.5+3.5 case and the 3+3+1 case -- in order to gauge the benefits of running LBNF in the so-called high energy configuration. In some cases, see, for example, Fig.~\ref{fig:NonUnitary}, the larger $\nu_{\tau}$-appearance sample leads to better sensitivity, while in other cases the two running strategies are practically identical. Overall, our study does not unambiguously indicate that running part of the time in high-energy mode is qualitatively better than running in the so-called CP-optimized mode~\cite{LauraFlux,Acciarri:2015uup}.  

While there is a guarantee flux of $\nu_{\tau}$ in the DUNE far detector, one can also search for $\nu_{\tau}$ in the DUNE near detector. In this case, the neutrino flux -- combining all flavors -- is huge, while the beam-related backgrounds are zero: virtually no $\nu_{\tau}$ are produced as a direct consequence of the original beam--target collision. This allows one to perform ``background free'' searches for $\nu_{\tau}$-appearance and, we expect, probe some of the scenarios discussed here with great sensitivity. Similar searches are currently being pursued by the NO$\nu$A collaboration \cite{ Keloth:2017vdp}. This subject is beyond the aspirations of this manuscript. When it comes to $\nu_{\tau}$-appearance, there are also neutrino sources for the DUNE far detector other than the LBNF beam. The atmospheric $\nu_{\tau}$ flux is large and spans a few energy decades. Some of the challenges of observing atmospheric $\nu_{\tau}$ in large liquid argon detectors, including the unknown neutrino direction and the absence of timing information, were explored in \cite{Conrad:2010mh}, and the subject is currently under investigation by the DUNE collaboration. 

\section*{Acknowledgements}
We thank Adam Aurisano and Jeremy Hewes for very instructive discussions and for information regarding the capabilities of the DUNE experimental setup. 
The work of AdG and KJK was supported in part by DOE grant \#de-sc0010143. The work of GVS was supported by FAPESP funding Grant  No. 2016/00272-9 and No. 2017/12904-2 and PP thanks the support of FAPESP-CAPES funding
grant 2014/05133-1, 2014/19164-6 and 2015/16809-9 Also the partial support from FAEPEX funding grant, No 2391/17 and the Fermilab NPC grant. GVS and PSP also thanks the partial support of the Coordena\c{c}\~ao de Aperfei\c{c}oamento de Pessoal de N\'ivel Superior - Brasil (CAPES) - Finance Code 001. This manuscript has been authored in part by Fermi Research Alliance, LLC under Contract No. DE-AC02-07CH11359 with the U.S. Department of Energy, Office of Science, Office of High Energy Physics.

\appendix

\section{The $\nu_{\mu}$ and $\nu_e$ Samples at DUNE}
\label{appendix:others}

Here we present the details of our simulations for the $\nu_\mu \to \nu_e$ appearance and $\nu_\mu \to \nu_\mu$ disappearance channel analyses, updated from Refs.~\cite{Berryman:2015nua,deGouvea:2015ndi,Berryman:2016szd,deGouvea:2016pom,deGouvea:2017yvn}. As in the case of the expected $\nu_{\tau}$-yields, discussed in Sec.~\ref{sec:DUNE}, the LBNF-DUNE neutrino and antineutrino fluxes match those available in Refs.~\cite{LauraFlux,Acciarri:2015uup} and when computing the effects of neutrino oscillations we have used the oscillation parameters listed in Eq.~(\ref{eq:nufit}).

\begin{figure}[ht]
\begin{center}
\includegraphics[width=\textwidth]{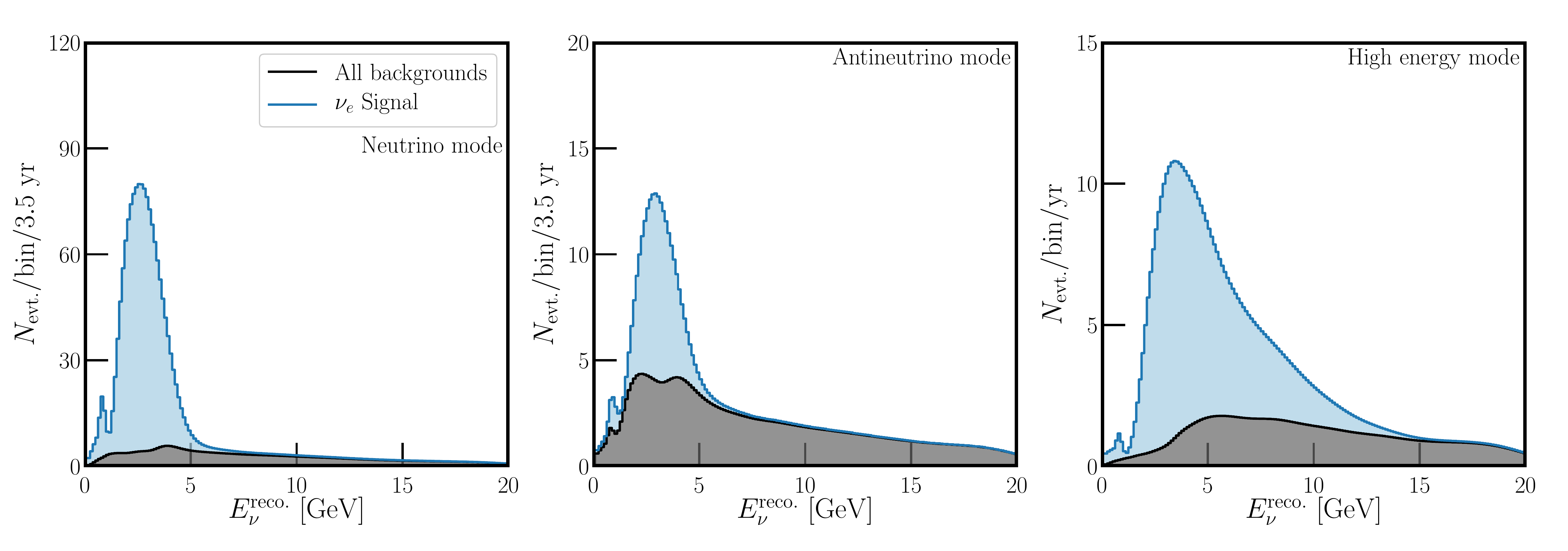}
\caption{Expected number of $\nu_\mu \to \nu_e$ signal events (blue) and background events (black). See text for details. The left panel displays the number of expected events in each $0.125$ GeV bin for the neutrino beam mode assuming 3.5 years of data collection, the center panel displays antineutrino mode yields, and the right panel displays high-energy mode yields, assuming 1 year of data collection.}
\label{fig:AppYields}
\end{center}
\end{figure}

Fig.~\ref{fig:AppYields} depicts signal (blue) and background (black) event yields for the $\nu_\mu \to \nu_e$ appearance channel as a function of the reconstructed neutrino energy, assuming 3.5 years of data collection in neutrino mode (left), 3.5 years antineutrino mode (center), and 1 year of high-energy mode (right). The data are organized into energy bins of width $0.125$ GeV. The signal events are stacked on top of backgrounds, which consist of the following: opposite-sign background ($\bar{\nu}_\mu \to \bar{\nu}_e$); intrinsic $\nu_e$ beam-contamination; misidentified $\nu_\mu$ charged-current events where a muon is misidentified as an electron; misidentified $\nu_\tau$ charged-current events; and neutral-current events, misidentified as $\nu_e$ charged-current events.

\begin{figure}[ht]
\begin{center}
\includegraphics[width=\textwidth]{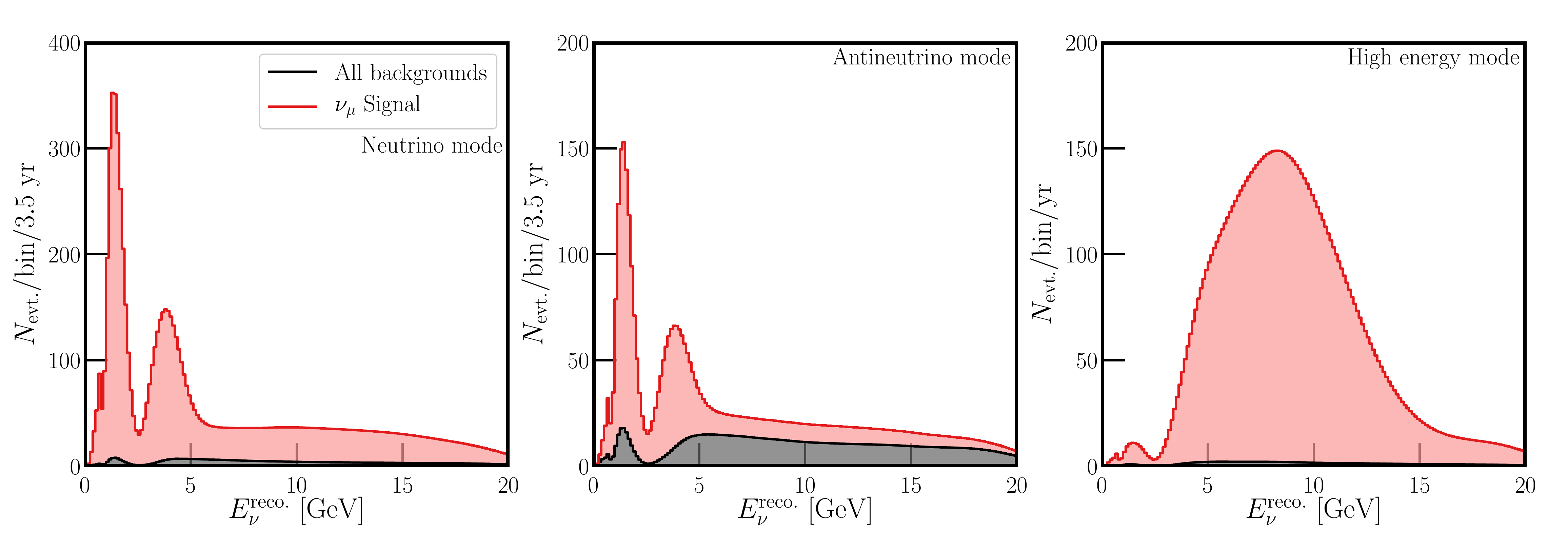}
\caption{Expected number of $\nu_\mu \to \nu_\mu$ signal events (red) and background events (black). See text for details. The left panel displays the number of expected events in each $0.125$ GeV bin for the neutrino beam mode assuming 3.5 years of data collection, the center panel displays antineutrino mode yields, and the right panel displays high-energy mode yields, assuming 1 year of data collection.}\label{fig:DisYields}
\end{center}
\end{figure}

Likewise, in Fig.~\ref{fig:DisYields} depicts the expected signal (red) and background (black) event yields for the $\nu_\mu \to \nu_\mu$ disappearance channel for the same beam configurations as in Fig.~\ref{fig:AppYields}. The data are also organized into energy bins of width $0.125$ GeV. Here, the backgrounds considered are the opposite-sign background ($\bar{\nu}_\mu \to \bar{\nu}_\mu$); misidentified $\nu_\tau$ charged-current events where the $\tau$ is misidentified as $\mu$; and neutral-current events misindentified as muon charged-current events.

The distributions in Figs.~\ref{fig:AppYields} and~\ref{fig:DisYields} are as a function of reconstructed neutrino energy: we consider an energy resolution of $\sigma_E = 7\% \left(\frac{E_\nu}{1\ \mathrm{GeV}}\right) + 3.5\%\sqrt{\frac{E_\nu}{1\ \mathrm{GeV}}}$, consistent with the goals of the DUNE experiment~\cite{Acciarri:2015uup}. When performing analyses including the appearance and disappearance channels, we include a normalization uncertainty of $5\%$ as a nuisance parameter, marginalized over when presenting results.

\section{More on the $\nu_{\mu}\to\nu_{\tau}$ Oscillation Probability at DUNE}
\label{appendix:prob}

When computing neutrino oscillation probabilities in constant matter, a very good approximation for the LBNF-DUNE experimental setup, one must contend with three different oscillation frequencies: 
\begin{equation}
\Delta_{31}\equiv\Delta m^2_{31}/2E, ~~~ \Delta_{21}\equiv\Delta m^2_{21}/2E, ~~~ A\equiv\sqrt{2}G_Fn_e, 
\end{equation}
where $n_e$ is the electron number density in the medium. For LBNF-DUNE, $A\sim 5.8\times 10^{-4}$~eV$^2/$GeV such that both $\Delta_{31}$ and $A$ are much larger than $\Delta_{21}<10^{-5}$~eV$^2/$GeV for neutrino energies above $\tau$-threshold. Moreover, when it comes to $\nu_{\mu}\to\nu_{\tau}$ oscillations, all relevant elements of the mixing matrix -- $U_{\alpha i}$, for $\alpha=\mu,\tau$ and $i-1,2,3$ are of the same order of magnitude. These two facts imply that, for neutrino energies about $\tau$-threshold, $P(\nu_{\mu}\to\nu_{\tau})$ depends very weakly on ``solar'' effects and setting  $\Delta m^2_{21}$ to zero is a good approximation. 

In the limit $\Delta m^2_{21}\to 0$, $P(\nu_{\mu}\to\nu_{\tau})$ can be computed analytically and expressed as
\begin{equation}
P(\nu_{\mu}\to\nu_{\tau}) =  \sin^22\theta_{23}\left|e^{\left(i\frac{\Delta_ML}{2}\right)}\cos^2\theta_M\sin\left(\frac{\Delta_+L}{2}\right)+\sin^2\theta_M\sin\left(\frac{\Delta_-L}{2}\right)\right|^2,
\end{equation}
where
\begin{eqnarray}
\Delta_{\pm}&=&\frac{A+\Delta_{31}}{2}\pm\frac{\Delta_M}{2}, \\
\Delta_M&=&\sqrt{\left(A-\Delta_{31}\cos2\theta_{13}\right)^2+\Delta_{31}^2\sin^22\theta_{13}}, \\
\Delta_M\sin2\theta_{M}&=&\Delta_{31}\sin2\theta_{13}, \\
\Delta_M\cos2\theta_{M}&=&\Delta_{31}\cos2\theta_{13}-A.
\end{eqnarray}
This expression, it turns out, is very well approximated by the vacuum one, Eq.~(\ref{eq:pmutau}) for the neutrino energies and baseline of interest. Numerically, 
\begin{equation}
\Delta_{31}=3.6\times 10^{-4}~\frac{\rm eV^2}{\rm GeV}\left(\frac{3.5~\rm GeV}{E}\right)\left(\frac{\Delta m^2_{31}}{2.5\times 10^{-3}~\rm eV^2}\right),
\end{equation}
and hence $A>\Delta_{31}$ for neutrino energies above $\tau$-threshold. It is easy to show that
\begin{equation}
P(\nu_{\mu}\to\nu_{\tau})=\sin^22\theta_{23}\left|\sin\left[\frac{\Delta_{31}L}{2}\left(\cos^2\theta_{13}+{\cal O}\left(\frac{\Delta_{31}}{A}\right)\right)\right]+{\cal O}\left(\frac{\Delta_{31}}{A}\right)\right|^2.
\end{equation}
In the limit $A\gg \Delta_{31}$ (when $\Delta_{31}L\ll 1$, see Eq.~(\ref{eq:phase})), $P(\nu_{\mu}\to\nu_{\tau})$ in matter is very well approximated by the vacuum expression, Eq.~(\ref{eq:pmutau}). We have verified numerically that matter effects are also negligible close to $\tau$-threshold, thanks in no small part to the fact that $\sin^2\theta_{13}$ is very small.

The inclusion of solar effects -- i.e., including nonzero $\Delta m^2_{21}$ -- is much more cumbersome -- see for example, \cite{Denton:2016wmg} -- and does not modify the conclusions reached here, as we verified both semi-analytically, using the results of \cite{Denton:2016wmg}, and numerically. 

\bibliographystyle{kp}
\bibliography{tau_bib}{}

\end{document}